%% file: main.tex
\documentclass[acmsmall,screen,dvipsnames,nonacm]{acmart}

\input{preamble/imports.tex}
\input{preamble/alias.tex}



\setcopyright{cc}
\setcctype{by}
\acmDOI{10.1145/3798203}
\acmYear{2026}
\acmJournal{PACMPL}
\acmVolume{10}
\acmNumber{OOPSLA1}
\acmArticle{95}
\acmMonth{4}
\received{2025-10-08}
\received[accepted]{2026-02-17}

\begin{document}
\definecolor{cbpink}{RGB}{203,107,135}

\title{Floating-Point Usage on GitHub: A Large-Scale Study of Statically Typed Languages}

\author{Andrea Gilot}
\orcid{0009-0006-4463-9414}
\affiliation{%
  \institution{Uppsala University}
  \city{Uppsala}
  \country{Sweden}
}
\email{andrea.gilot@it.uu.se}

\author{Tobias Wrigstad}
\orcid{0000-0002-4269-5408}
\affiliation{%
  \institution{Uppsala University}
  \city{Uppsala}
  \country{Sweden}
}
\email{tobias.wrigstad@it.uu.se}

\author{Eva Darulova}
\orcid{0000-0002-6848-3163}
\affiliation{%
  \institution{Uppsala University}
  \city{Uppsala}
  \country{Sweden}
}
\email{eva.darulova@it.uu.se}

\begin{abstract}

    Reasoning about floating-point arithmetic is notoriously hard. While static and dynamic analysis techniques or program repair have made significant progress, more work is still needed to make them relevant to real-world code. On the critical path to that goal is understanding what real-world floating-point code looks like.

    To close that knowledge gap, this paper presents the first large-scale empirical study of floating-point arithmetic usage across public GitHub repositories. We focus on statically typed languages to allow our study to scale to millions of repositories. We follow state-of the art mining practices including random sampling and filtering based on only intrinsic properties to avoid bias, and identify floating-point usage by searching for keywords in the source code, and programming language constructs (\eg{} loops) by parsing the code. Our evaluation supports the claim often made in papers that floating-point arithmetic is widely used. Comparing statistics such as size and usage of certain constructs and functions, we find that benchmarks used in literature to evaluate automated reasoning techniques for floating-point arithmetic are in certain aspects representative of `real-world' code, but not in all.

    We publish a dataset of 10 million real-world floating-point functions extracted from our study. We demonstrate in a case study how it may be used to identify new floating-point benchmarks and help future techniques for floating-point arithmetic to be designed and evaluated to match actual users’ expectations.

\end{abstract}

\maketitle

\input{sections/introduction.tex}
\input{sections/related.tex}

\input{sections/methodology.tex}


\input{sections/data_collection.tex}

\input{sections/results.tex}

\input{sections/case_study.tex}
\input{sections/discussion.tex}

\section{Data-Availability Statement}


All code, datasets, and artifacts necessary to reproduce the analyses and results presented in this paper are publicly available and permanently archived.
\begin{itemize}
    \item \textit{Scyros.} The source code of Scyros, our framework for large-scale code studies, is available on GitHub under the Apache 2.0 license at: \url{https://github.com/fxpl/scyros}. A versioned snapshot corresponding to the evaluated artifact is archived on Zenodo along with the artifact.
    \item \textit{Function Dataset.} The corpus of metadata for floating-point functions extracted from GitHub, including links to the original functions, is archived on Zenodo under the Apache License 2.0 at: \url{https://doi.org/10.5281/zenodo.17055622}.
    \item \textit{C Challenge Benchmarks:} the set of 59~self-contained C floating-point benchmarks extracted from our corpus is available on Zenodo as part of the function dataset.
    \item \textit{Floating-Point Keywords List:} the list of keywords we used to identify floating-point code is also available on Zenodo as part of the function dataset.
    \item \textit{Artifact.} 
    The artifact accompanying this paper is a local web application that reproduces all analyses and results reported in the paper and provides a step-by-step guide for reproducing the study with different keywords on a small set of newly mined repositories. The artifact is archived on Zenodo under the Apache 2.0 license at \url{https://doi.org/10.5281/zenodo.18500268} and has been submitted for Artifact Evaluation \cite{gilot_2026_18500269}.
    
\end{itemize}

\section*{Acknowledgments}
We thank the reviewers for their thoughtful feedback that strengthened our work. Additionally, we would like to thank 
Fridtjof Stoldt, Noé De Santo and Paulo Canelas for their suggestions and feedback on this work.

This work is co-funded by the European Union (ERC, HORNET, 101163629).
Views and opinions expressed are however those of the author(s) only and do not
necessarily reflect those of the European Union or the European Research Council.
Neither the European Union nor the granting authority can be held responsible for them.

This research was supported by the \grantsponsor{VR}{Swedish
Research Council}{https://vr.se} through the grant
\grantnum{VR}{2023-05161}, ``Scalable Error Verification: How
Accurate is your Numerical Code?'', and \grantnum{VR}{2024-04565}, ``Data-race Freedom and Memory Safety for Untyped Languages''.

\bibliographystyle{ACM-Reference-Format}
\bibliography{bibliography/floating_point_verification_clean, bibliography/github_mining_clean}

\end{document}

%% file: preamble/imports.tex
\usepackage{amsmath}
\usepackage{pgfplots}
\usepackage{pgfplotstable}
\pgfplotsset{compat=1.17}
\usepackage{ragged2e}
\usepackage[most]{tcolorbox}
\usepackage{hyperref}
\usepackage[capitalise,nameinlink]{cleveref}
\usepackage{subcaption}
\usepackage{tabularx}
\usepackage{multirow}
\usepackage{adjustbox}
\usepackage{colortbl}
\usepackage{threeparttable}
\usepackage{enumitem}
\usepackage{soul}
\usepackage{wrapfig}
\usepackage{siunitx}

\usepackage{siunitx}
\usepackage{tikz}
\usetikzlibrary{shapes.geometric, positioning, patterns}

\usepackage{booktabs}
\usepackage{xspace}

\usepackage{listings}
\usepackage[T1]{fontenc}  
\usepackage[scaled=0.75]{beramono}  

%% file: preamble/alias.tex
\newcommand{\agethreshold}[0]{2}
\newcommand{\statlang}[0]{51}

\newcommand{\days}{\text{d}}

\newcommand{\minutes}{\text{min}}

\newcommand{\millions}{\text{M}}
\newcommand{\thousands}{\text{k}}

\newcommand{\Pcent}[1]{#1\,\%}

\newcommand{\ts}[0]{TypeScript}

\definecolor{cbred}{HTML}{D55E00}
\definecolor{cborange}{HTML}{E69F00}
\definecolor{cbyellow}{HTML}{F5C710}
\definecolor{cbgreen}{HTML}{009E73}
\definecolor{cbcyan}{HTML}{56B4E9}
\definecolor{cbblue}{HTML}{0072B2}
\definecolor{cbpurple}{HTML}{CC79A7}

\definecolor{darkblue}{HTML}{00008B}
\definecolor{gostring}{RGB}{150,40,40}
\lstdefinestyle{highlight}{
  basicstyle=\ttfamily\small,
  columns=fullflexible,
  breaklines=true,
  showstringspaces=false,
  tabsize=2,
  keywordstyle=\color{blue}\bfseries,
  stringstyle=\color{gostring},
  commentstyle=\color{cbgreen}\itshape,
  identifierstyle=\color{black},
}

\newcommand{\sideboxcolorprimary}{MidnightBlue!80!black}
\newcommand{\sideboxcolorsecondary}{MidnightBlue!60!white}

\newtcolorbox{rqanswer}[1][]{
    enhanced,
    breakable,
    colback=gray!10!white,
    colframe=gray!50!black,
    boxrule=1.5pt,
    arc=2mm,
    left=4pt,
    right=4pt,
    top=4pt,
    bottom=4pt
}

\newtcolorbox{sidebox}[2][]{
    enhanced,
    breakable,
    colback=white, 
    colframe=\sideboxcolorsecondary, 
    fontupper=\ttfamily, 
    boxrule=1.5pt, 
    arc=3mm,
    left=3mm, 
    right=3mm,
    top=3mm,
    bottom=2mm,
    boxed title style={
      colback=\sideboxcolorprimary,
      colframe=\sideboxcolorprimary,
      fonttitle=\bfseries,
      size=small,
      arc=2.5mm, 
    },
    attach boxed title to top left={ 
      yshift=-3mm,
      xshift=3mm
    },
    title={#2},
}

\newcommand{\eg}{\emph{e.g.,}}
\newcommand{\ie}{\emph{i.e.,}}

\newcommand{\code}[1]{\lstinline[basicstyle=\fontsize{10}{10}\selectfont\tt,keywordstyle=\fontsize{10}{10}\selectfont\bf]@#1@}

\newcommand{\ku}[1]{\num{#1}\,k}

\newif\ifdiff
 \difffalse

%% file: sections/introduction.tex
\section{Introduction}

Numerical software is widely used across different domains such as embedded
systems, scientific computing, and machine learning. Such software frequently makes use of
floating-point arithmetic to efficiently approximate computations with real
numbers. Some approximation of the infinitely precise reals is fundamentally and practically necessary, and
introduces rounding errors at most arithmetic operations as well as potentially
the special values infinity and Not-a-Number (NaN). Overall this makes reasoning about
the correctness of such software unintuitive, error prone and time consuming.

This could be the beginning of a paper on automated reasoning for
floating-point programs, \eg{} with a
static~\cite{darulovaDaisyFrameworkAnalysis2018,solovyevRigorousEstimationFloatingPoint2018}
or a dynamic
approach~\cite{zouDetectingFloatingpointErrors2019,guoEfficientGenerationErrorinducing2020,zouOraclefreeRepairSynthesis2022}.
Such an approach would typically be evaluated either using a small benchmark suite~\cite{damoucheStandardBenchmarkFormat2017}, or on a relatively small set
of benchmarks hand-picked specifically for that paper.

While hand-picked benchmarks can be helpful to demonstrate particular features
and issues, it is unclear how they generalise to ``real-world'' code. To the
best of our knowledge, no large-scale code study has investigated how representative
the currently used floating-point benchmarks are of ``real-world'' usage, \eg{} in terms of
size, operations performed, or frequency of library function usage.
Neither do we know how prevalent the use of floating-point arithmetic is in modern software.

While real-world floating-point usage is anecdotal, the difficulty in
developing automated reasoning techniques for floating-point arithmetic is well
documented. For example, static analysis techniques bounding worst-case
rounding errors~\cite{titoloRigorousFloatingPointRoundOff2025,darulovaDaisyFrameworkAnalysis2018,solovyevRigorousEstimationFloatingPoint2018} are limited to small programs (mostly individual
functions) and have no or limited support for common program constructs such as
conditionals and loops. This is also reflected in the standard benchmark set~\cite{damoucheStandardBenchmarkFormat2017}
used in this domain.
Dynamic approaches aiming to identify specific inputs that induce large rounding
errors~\cite{guoEfficientGenerationErrorinducing2020,zouDetectingFloatingpointErrors2019},
or for repairing such errors~\cite{zouOraclefreeRepairSynthesis2022,panchekhaAutomaticallyImprovingAccuracy2015} are
evaluated on different, hand-picked sets of benchmarks, most of which are
typically small (\eg{} individual functions).
In summary, the current benchmarks used
align with the current capabilities of the tools,
but it is unclear to what extent they align with actual users' expectations
in terms of the code they want the tools to reason about. While
community benchmarks are a great way of focusing research, unless
the benchmarks are ``relevant'', they may end up directing research in
the wrong direction.

This paper presents the first \emph{large-scale} study of real-world floating-point
code. The aim is to inform current and future research efforts in this domain,
for example by aiding (more) representative selections of benchmarks for
evaluation, and to guide the development of future benchmarks and reasoning techniques in a practically relevant direction.
With that in mind, we address the following research questions:
\begin{itemize}[label={}, itemindent=0ex, leftmargin=0ex, itemsep=1ex]
    \item RQ1. How prevalent is floating-point arithmetic in open-source statically typed code on GitHub?
    \item RQ2. What are the characteristics of floating-point code in terms of programming language constructs and size?
    \item RQ3. How representative are benchmarks used in the literature to evaluate floating-point analyses of real-world floating-point code?
\end{itemize}

Characterising ``real-world floating-point code'' is non-trivial for several reasons.
First, limiting ourselves to publicly available code,
real-world code bases use a multitude of programming languages and the volume is
too large to be processed and inspected in full. The latter can be addressed
by sampling, but 
to avoid measurement bias, the code must be checked for code
clones~\cite{lopesDejaVuMapCode2017}, and filtered to exclude projects that
do not contain any code at all and ``toy'' projects (\eg{} projects that were never updated after the initial commit).
While a convenient and continuously updated dataset of GitHub projects~\cite{dabicSamplingProjectsGitHub2021} aimed
for code studies as ours exists, it includes only
projects with at least ten stars (to reduce the amount of data). Such filtering
has been shown to skew the results significantly~\cite{majFaultOurStars2024},
making it unsuitable to generalise from.

Second, once we have a set of ``interesting'' projects, there is no existing
tool that can efficiently \emph{identify} and \emph{analyse} floating-point
code. To obtain representative results, such an analysis must support a multitude of programming
languages and be automated to be able to handle sufficiently large volumes of code.
Our insight is that in \emph{statically typed} languages the type annotations on
\eg{} function parameters allow us to identify floating-point code with an
inexpensive and largely programming language-agnostic approach by ``grepping''
for suitable keywords in source code.
To identify floating-point code, \emph{type} annotations are critical; it is not enough
to look for the presence of arithmetic operators (\eg{} \code{+}) as these are often
overloaded to work for integers, strings or other types.
This is unlike other code properties such as imports of specific libraries that
can be unambiguously identified even in dynamically typed source
code.\footnote{Our study shows that mathematical library functions (\eg{}
\texttt{sin}, \texttt{exp}) cannot be used to identify floating-point code either, as much
floating-point code does not use them.} In this paper, we therefore limit
ourselves to studying floating-point usage in statically typed languages.



\paragraph*{Contributions}
This paper makes the following contributions:

\paragraph{\textbf{Large-Scale Floating-Point Code Mining}}
We design and implement a large-scale mining methodology
(\Cref{sec:methodology}) for detecting and analysing floating-point usage in
statically typed languages at the project, file and function level. Our approach
is fully automated and combines random sampling, project filtering based on intrinsic properties, and
fuzzy duplicate elimination to produce a reliable dataset.

Our methodology identifies floating-point usage in statically typed programming
languages using \emph{keywords} in the source code directly. This allows us to consider
\emph{all} statically typed languages that are recognised by GitHub in a scalable and fully
automated way, as long as these languages have floating-point types and a language reference.
For the most commonly used languages (covering more than \Pcent{92} of the sampled
code), we parse the code to identify programming language constructs (\eg{}
loops) using a language-agnostic parsing library.

Large parts of the methodology and implementation are not
specific to floating-point analysis (nor to statically typed code, although we rely on types for identifying relevant code) and we expect them to be reusable in other
large-scale empirical code studies.
The code is available as open-source\footnote{Available at: \url{https://github.com/fxpl/scyros}.}.

\paragraph{\textbf{Quantitative Analysis}}
We analyse \num{447 209} (randomly sampled) GitHub projects (\Cref{sec:data-collection}) and present the results in \Cref{sec:results}.
Our analysis shows that, with \Pcent{95} confidence, over \Pcent{62} of the projects contain floating-point code,
supporting the often made claim that floating-point numbers are widely used.
While this confirms folklore beliefs, we are the first to draw this conclusion based on a large-scale code study of open-source code.

Most functions in our final dataset tend to be small, and function calls and
conditional statements appear more often than loops and (explicitly mentioned)
special values.
The FPBench~\cite{damoucheStandardBenchmarkFormat2017} benchmark suite used in literature to evaluate floating-point techniques
shows different characteristics, \eg{} involving fewer conditionals and more
transcendental library function calls. 
Functions from the GNU Scientific Library, also sometimes used in evaluations, appear rarely in our dataset.

To the best of our knowledge, our study is the first to provide a clear picture
of how floating-point numbers are actually used in real-world code, and
indicates that currently used floating-point benchmarks are largely not
representative of such code.

\paragraph{\textbf{Floating-Point Functions Dataset and Challenge Benchmarks}}
For the benefit of other researchers,
we release a dataset of 10 million real-world floating-point functions extracted from our study%
\footnote{Available at \url{https://doi.org/10.5281/zenodo.17055622}.}.
This dataset focuses on functions, since most static and dynamic
floating-point reasoning tools today work on a per-function basis, so this dataset
is most (immediately) relevant for their benchmarking.
We intend this dataset to serve as a foundation for constructing new realistic
benchmarks and evaluating floating-point reasoning tools in a practically relevant way,
and invite others to contribute in this undertaking.

We show how our corpus can be used to generate realistic, ready-to-run benchmarks that reflect real-world floating-point practice via a case study (\Cref{sec:case-study}) that extracts 59 self-contained C floating-point benchmarks from functions in our dataset.

%% file: sections/related.tex
\section{Related Work}

\paragraph{\textit{Floating-Point Reasoning Benchmarks}} Most works that focus on numerical programs are evaluated using different sets of
hand-picked programs. Hand-picking programs can be good for stressing certain
types of problems, but can just as easily---intentionally or not---suppress other
problems and challenges. It is also unclear how to generalise from hand-picked
problems to code ``in the real world''. (Indeed, this was part of what prompted this research
in the first place.)

We are aware of only a single benchmark suite
created specifically for floating-point analysis:
FPBench~\cite{damoucheStandardBenchmarkFormat2017}. FPBench
is typically used to evaluate static analysis tools~\cite{titoloRigorousFloatingPointRoundOff2025}.
The FPBench benchmarks were collected from individual papers~\cite{darulovaDaisyFrameworkAnalysis2018,solovyevRigorousEstimationFloatingPoint2018} where they were hand-picked and are typically isolated arithmetic expressions.
They are therefore not likely representative of challenges found in real-world programs. Techniques that
target programs outside of this set choose their own set of benchmarks, \eg{}
programs that are larger~\cite{dasScalableFloatErrorAnalysis2020} or that use
specific data structures~\cite{isychevScalingRoundingAnalysisFunctionalDS2023}.

Some techniques are evaluated on a small number of hand-picked functions from numerical libraries, \eg{} from the GNU
Scientific Library (GSL)~\cite{miaoCompilerNumericalInconsistencies2024,
guoEfficientGenerationErrorinducing2020, zouDetectingFloatingpointErrors2019,
zouOraclefreeRepairSynthesis2022}, occasionally extended by some additional functions.
For example, Di Franco et al.~\cite{difrancoNumericalBugCharacteristics2017} study bug characteristics
in numerical libraries (NumPy, LAPACK, GSL, etc.), and
Liew et al.~\cite{liewFloatSymbolicExecution2017} have two teams independently hand-pick benchmarks for evaluating
symbolic execution.
In the HPC community, there are examples of picking a specific proxy application, such as Lulesh,
as a representative of ``HPC code''~\cite{wangPrecisionTuningGNN2024}. The tool NSan~\cite{courbetNSanFloatingpointNumerical2021} is evaluated on the commercial SPECfp2006
benchmark set that contains scientific computing applications and that was designed for performance benchmarking.
In what way these selected benchmarks are representative of general real-world code,
\eg{} beyond just library code or a specific application domain, is unclear and not discussed in these papers.

\paragraph{\textit{Real-World Usage through Code Studies}} We are not aware of any large-scale code study that investigates characteristics of real-world floating-point code.
Previous works have investigated how different features are used
in practice in different programming languages, for example in the context of static analysis for the R programming
language~\cite{sihlerAnatomyRealWorldCode2024},
dynamic features in Smalltalk~\cite{callauHowDevelopersUseDynamic2011},
the use of \code{eval} and dynamic behaviour in JavaScript~\cite{richardsTheEvalThatMenDo2011,richardsAnAnalysisOfDynamicBehaviour2010},
dynamic features or code complexity in Python~\cite{akerblomTracingDynamicFeatures2014,grotovComparisonPythonJupyter2022},
inheritance, method chaining, and streams in Java~\cite{temperoHowDoJavaProgramsUseInheritance2008,dyerMiningBillionsAST2014,khatchadourian2020,keshkMethodChainingRedux2023}, or
gradual types in TypeScript and Python~\cite{troppmannTypedConfused2024}.
While this paper also studies features of code, these are sufficiently different
that we cannot straightforwardly re-use existing infrastructure. We furthermore
consider floating-point usage across many different programming languages.

While some features in dynamically typed programming languages such as Python
can be detected statically~\cite{pengEmpiricalStudyCommon2021}, others---specifically
types---require programs to be run on suitable
inputs~\cite{akerblomTracingDynamicFeatures2014}. For example, to determine whether
an addition (\code{+}) operates over floating-point values requires detection of the
operand types at runtime.
This cannot be straightforwardly automated; to run code it is necessary to
obtain---for each project individually---all its required dependencies, and
understand how to run the program in a representative way (including representative
and meaningful input data and GUI interactions). Such information is not
available in a consistent format that can be automatically processed at a large
scale. We thus focus on statically typed programming languages only in this
study.

\paragraph{\textit{GitHub Mining}}
Our methodology draws inspiration from state-of-the-art practices in GitHub mining.
Prior studies show that the majority of GitHub repositories are inactive, personal, or not software projects at all, and that most of the files are duplicates of others \cite{kalliamvakouPromisesPerilsMining2014,lopesDejaVuMapCode2017, jupyterNotebooksonGitHub}.
Other work demonstrates that filtering repositories using extrinsic characteristics such as the number of stars introduces systematic biases in software repository studies \cite{borgesWhatsGitHubStar2018,majFaultOurStars2024}.
In a reproduction study, Berger et al. reveal the risks of ``careless'' keyword searches (\eg{} tagging commits as bug fixes if they contain the word \texttt{"fix"} without excluding \texttt{"infix"}, \texttt{"suffixes"}, or \texttt{"does not fix"} etc.) and emphasise  the necessity of manual validation to ensure methodological soundness \cite{berger_impact_2019}.

Several frameworks exist to simplify code studies on GitHub, but none fully satisfy our requirements.
Queryable datasets mined from GitHub exist, yet they are either outdated \cite{dyerBoaUltraLargeScale2015}, affected by inconsistencies~\cite{gousios2012ghtorrent},
or rely on GitHub stars \cite{dabicSamplingProjectsGitHub2021,dabicSeartDataHub2024} that have been shown to bias results~\cite{majFaultOurStars2024}.
World of Code~\cite{maWorldCodeEnabling2021} continuously aggregates repositories from multiple sources and could be used to implement parts of our methodology,
but it does not expose part of the metadata we require and does not allow downloading of full projects.
Other frameworks target different aspects of repository analysis.
CodeDJ~\cite{majCodeDJReproducibleQueries2021} ensures reproducible queries over projects by downloading full project histories, which exceeds our needs and would lead to an excessive amount of data being downloaded; the analysis would also still need to be implemented ``from scratch''.
GitCProc~\cite{casalnuovoGitcprocToolProcessing2017} enables keyword searches in commit histories but supports only a limited set of languages and does 
not exclude comments or string literals, which compromise validity.
CodeQL~\cite{githubcodeql_2025} offers semantic analysis across multiple languages but requires full-project access, raising scalability concerns.

%% file: sections/methodology.tex
\section{Methodology}\label{sec:methodology}

Our goal is to study the prevalence and type of code operating on floating-point
numbers in statically typed ``real-world'' code. To that end, we design a methodology that mainly
follows the approach suggested by Maj et al. \cite{majFaultOurStars2024} and
that is designed in accordance with the ACM SIGSOFT Empirical Standards for
repository mining \cite{ralphEmpiricalStandardsSoftware2021}. We first give an
overview here, before discussing details in the next subsections.

\input{sections/overview.tex}

We now expand on the details of each step of the pipeline and choices made during the design of the methodology.

\subsection{Project Selection}\label{sec:methodology-querying-metadata}

We sample ids of public repositories on GitHub (step~1) uniformly using GitHub's REST API. 
Given an id, the API returns the next 100 ids of public
repositories. To sample uniformly, we get the largest project id at the
beginning of data collection\footnote{At the time of writing, GitHub assigns
repository ids in chronological order. To obtain the current maximal id, we
create a new repository and query its id; the maximum id for our current dataset
was obtained on 24 January 2025.}, and generate a random integer between 0
and this maximal id to get the next following 100 ids.
The random number generator we use for sampling has a fixed seed to ensure reproducibility.
In addition to each project id, the API call returns the name of each repository and whether it is a fork.
Since we are sampling with replacement, we end up with duplicates that we remove after the sampling.
We also remove forks from the dataset as they bias the results of the study.


We collect metadata (step~2) for the sampled repositories using GitHub's REST API.
The metadata includes the size of the repository, its creation date, last push date, number of stars, number of issues, and its primary programming language.
As noted by Kalliamvakou et al.~\cite{kalliamvakouPromisesPerilsMining2014}, one of the challenges in mining GitHub is that a large portion of repositories are not software projects.
In addition, most of them are inactive and contain very few commits. 
Our target population consists of \emph{nontrivial} or ``non-toy'' software projects, but
defining what is nontrivial is difficult and subject to the authors'
interpretation.
As Maj et al.~\cite{majFaultOurStars2024} suggest, it is easier to discard
projects we know are \emph{not} part of the population than to define what is in
it.

We exclude a project from our dataset if any of the following hold (step~3):
\begin{itemize}[label={--}, itemindent=1ex, leftmargin=1em, itemsep=0ex]
\item it is smaller than \SI{50}{\kilo\byte},
\item GitHub does not assign a primary language to it (these are typically templates, or projects containing only documentation or binary files),
\item the creation date and last activity date are less than two months apart (so the "age" of the project is less than two months).
\end{itemize}
The rationale for using age as an exclusion criterion is to avoid including repositories in the population that were \eg{} used for student assignments.
An alternative or complement to the age is the number of commits, though in practice querying the number of commits using the API is expensive.

We do not filter based on social engagement, \ie{} stars or issue counts, as
it is unclear whether analysing only these repositories would generalise~\cite{majFaultOurStars2024,borgesWhatsGitHubStar2018}.
While the number of contributors could be used as a filter, we choose not to as single-maintainer projects could still reasonably be part of the population.

\subsection{Language Filtering}

Our target population consists of code written in statically typed programming languages. 
Since the metadata we collect in the previous step includes only the primary language of each repository, we perform an additional query to retrieve the list of all the languages used in each project (step~4). 
As it is the last step before downloading repositories, we also query the hash of the latest commit on the main branch. 
This allows us to ensure the reproducibility of the results: if the repository remains reachable and its commit history has not been rewritten, it can be re-downloaded in the exact same state.

For each programming language recognised by GitHub, we determine whether it is
statically typed (based on its documentation) and, if so, include it in our list of target languages.
We exclude languages that do not have primitive floating-point types (\eg{} Coq, Agda, or Dafny).
We also exclude languages where we could not find the language reference (\eg{} Mirah or ActionScript).
This yields the list of \statlang{} statically typed programming languages in \Cref{fig:languages_list}. We discard repositories that do not contain code written in at least one of them (step~5).

\begin{figure}[t]
        \it Ada, AngelScript, Arduino, ATS, Ballerina, Beef, Boo, C, C\#, C++, Chapel,
        Clean, Crystal, Curry, CWeb, EC, Eiffel, Elm, F\#, F*, Fantom, Fortran,
        FreeBasic, Frege, Futhark, Gleam, Go, Haxe, Haskell, Java, Kotlin, Modula-2,
        Modula-3, Nemerle, Nim, Standard ML, Oberon, OCaml, Opa, Pascal, Reason, Rust, Purescript,
        Scala, Swift, Typescript, Vala, Uno, Volt, Xtend, and Zig.
        \caption{List of the \statlang{} 
          programming languages we include in this study.}
          \Description{List of statically typed programming languages analyzed, including C, C++, Java, Rust, Go, Swift, Scala, Kotlin, and Fortran.}
    \label{fig:languages_list}    
\end{figure}

\subsection{Downloading and Keyword Filtering}
We download (step~6) the content of the filtered repositories using the hash of the latest commit on the main branch we retrieved in the previous step.
We exclude files not written in a statically typed language. We determine programming language using file extensions,
as repositories do not contain metadata for mapping language to files.
We then compute the number of lines of code and the number of words.

We aim to retain files with functions operating on floating-point values. The
most precise approach to identify these would be to infer the type of each
individual operation, \ie{} run a type checker or type inference. This approach,
however, would require special treatment and significant effort for each
individual programming language.
We instead choose an approximate, but more scalable, approach, and aim to
identify functions with floating-point operations based on keywords used in the
source code via regular expressions.
Because the programming languages we are dealing with are all statically typed,
we know that function signatures often contain the types of the parameters and
return values. Hence, we search for floating-point types in the files using a
regular expression. A problem encountered by this approach is the presence of
type aliases, structures or classes that represent or contain floating-point
values and that might be defined in other files.
To mitigate this issue, we also look at other keywords related to floating-point computations, such as transcendental functions, rounding functions, and special values like NaN or infinity.
Looking at the presence of transcendental functions is also interesting per se, as they are a feature that several floating-point verification tools aim to support.
Finally, we examine the use of keywords related to arbitrary-precision floating-point computations to assess how frequently such features are used in practice.

For each of the 51 languages, we manually inspect the official language reference to identify floating-point types and special values, and we examine the corresponding standard math library to identify numerical functions.
We group keywords into four categories:
\begin{itemize}[label={--}, itemindent=1ex, leftmargin=1em, itemsep=0ex]
    \item Primitive floating point types: keywords used to declare standard floating point types, \eg{} \code{float}, \code{double}, \code{real}, \code{float32}, \code{float64}.
    \item Transcendental functions: keywords used to call transcendental functions, \eg{} \code{sin}, \code{cos}, \code{tan}, \code{exp}, \code{log10}.
    \item Other keywords related to floating point computations and special values, \eg{} \code{fma}, \code{posinf}, \code{nan}, \code{round}, \code{ulp}.
    \item Arbitrary precision floating-point functions and types, \eg{} \code{mpfr_t}, \code{bigfloat}, \code{apfloat}.
\end{itemize}
The JSON files containing the keyword definitions for each programming language are available along with our function dataset.

The search is case insensitive using boundaries such as spaces, punctuation or brackets to surround the words. 
While choosing keywords, we tried to be as exhaustive as possible while avoiding ambiguous keywords.
For example, keywords like \code{abs} or \code{pow} are valid floating-point functions
but they are often used in integer computations as well, and could lead to false positives.
While some keywords like \code{cos}, \code{sin} or \code{float} are common to most languages, others can be specific to a given programming language and ambiguous in others.
This is the case for the keyword \code{number} in Typescript, which may denote floating-points or integers, or \code{real} in Pascal or Fortran.
For each file, we count the number of keyword occurrences per category, and discard the file if it does not contain any keyword from neither of the four categories.


\subsection{De-Duplication}

Removing duplicates is a crucial step to avoid measurement bias,
as prior work shows that GitHub code contains considerable amounts of code duplication~\cite{lopesDejaVuMapCode2017}.
We eliminate files that are identical up to token reordering across repositories (step~7).
For this step and the following ones, we abandon the notion of repositories and treat files as standalone entities within a global pool. 
We represent each file as a bag-of-words, ignoring whitespace, delimiters, special characters, and token order. 
This representation enables the detection of token-level duplicates that differ only by whitespace changes or reordered declarations.
We compute the BLAKE3 hash of each bag-of-words representation and retain a single instance for each unique hash.

\subsection{Parsing and Function Extraction}

Once only deduplicated files containing numerical keywords are left, we parse them to extract individual functions (step~8).
We remove comments because they are not directly relevant to our analysis, may be outdated, and can bias results. 
In particular, terms such as double may appear in natural language text with meanings unrelated to floating-point types, 
thereby increasing the risk of false positives.
Comments are clearly valuable in their own right and deserve special
care, but we leave their analysis for future work.

We traverse the parse tree to extract individual functions.
For each function, we compute its metadata as the number of:
\begin{itemize}[label={--}, itemindent=1ex, leftmargin=1em, itemsep=0ex]
    \item lines of code;
    \item words\footnote{A word is a sequence of letters, numbers, or underscores delimited by boundaries (\eg{} brackets, punctuation, whitespaces, beginning and end of file).};
    \item parameters (\ie{} the function's arity);
    \item numerical types parameters;
    \item conditional statements in the body of the function, and their maximum nesting depth;
    \item loops in the body of the function, and their maximum nesting depth;
    \item function calls in the body of the function, and their maximum nesting depth (\eg{} \code{sin(cos(x))} has depth 2);
    \item occurrences of numerical keywords in the function.
\end{itemize}
We also record the function's location in the source file and whether a parsing error occurred, including its position when applicable.
Similar to comments, we exclude string literals from the keyword search, 
since keywords occurring in free text are more likely to produce false positives.
If a function contains at least one numerical keyword, we extract its code and store it in a separate file (step~9).

We parse the files using tree-sitter~\cite{Treesitter2025},
a language agnostic parsing library that already comes with grammars for the most
used programming languages. Our implementation currently supports C, C++, Java, TypeScript, C\# and Go,
which together account for more than \Pcent{92} of the functions collected. 
Supporting additional languages requires only defining the parse nodes of interest, 
which makes our implementation readily reusable and extendable in future code studies.
While other statistics, such as the number of floating-point arithmetic
operations, would have been relevant for the study, these rely on type
information, which is beyond tree-sitter's capacity\footnote{Tree-sitter
preserves type information that is explicitly declared in source code, but does
not do any type-checking, propagation or inference.}; \ie{} they would require
programming language-specific analyses.

%% file: sections/overview.tex
\subsection{Overview}\label{sec:overview}
We first define our \emph{target population}: functions from non-trivial projects
hosted on GitHub containing floating-point computations within code written in
statically typed languages.
GitHub alone contains an enormous amount of public repositories, and we assume
here that they form a representative sample of real world
software\footnote{Although our methodology and implementation use GitHub as the
primary data source, they are not specific to it. Both can be extended or
adapted to other hosting platforms.}.
We restrict our study to statically typed code so that we can automate the analysis
and thus to scale to significantly more projects.

Our final dataset consists of functions that use floating-point arithmetic.
We focus on individual functions as most floating-point
reasoning tools work on a per-function basis, though the corresponding files and
projects are easily identified. Our quantitative analysis considers
per-project, per-file and per-function properties, and so code not appearing in
functions will be considered in the first two.

\begin{figure*}[t]
    \centering
    \includegraphics[width=0.95\textwidth]{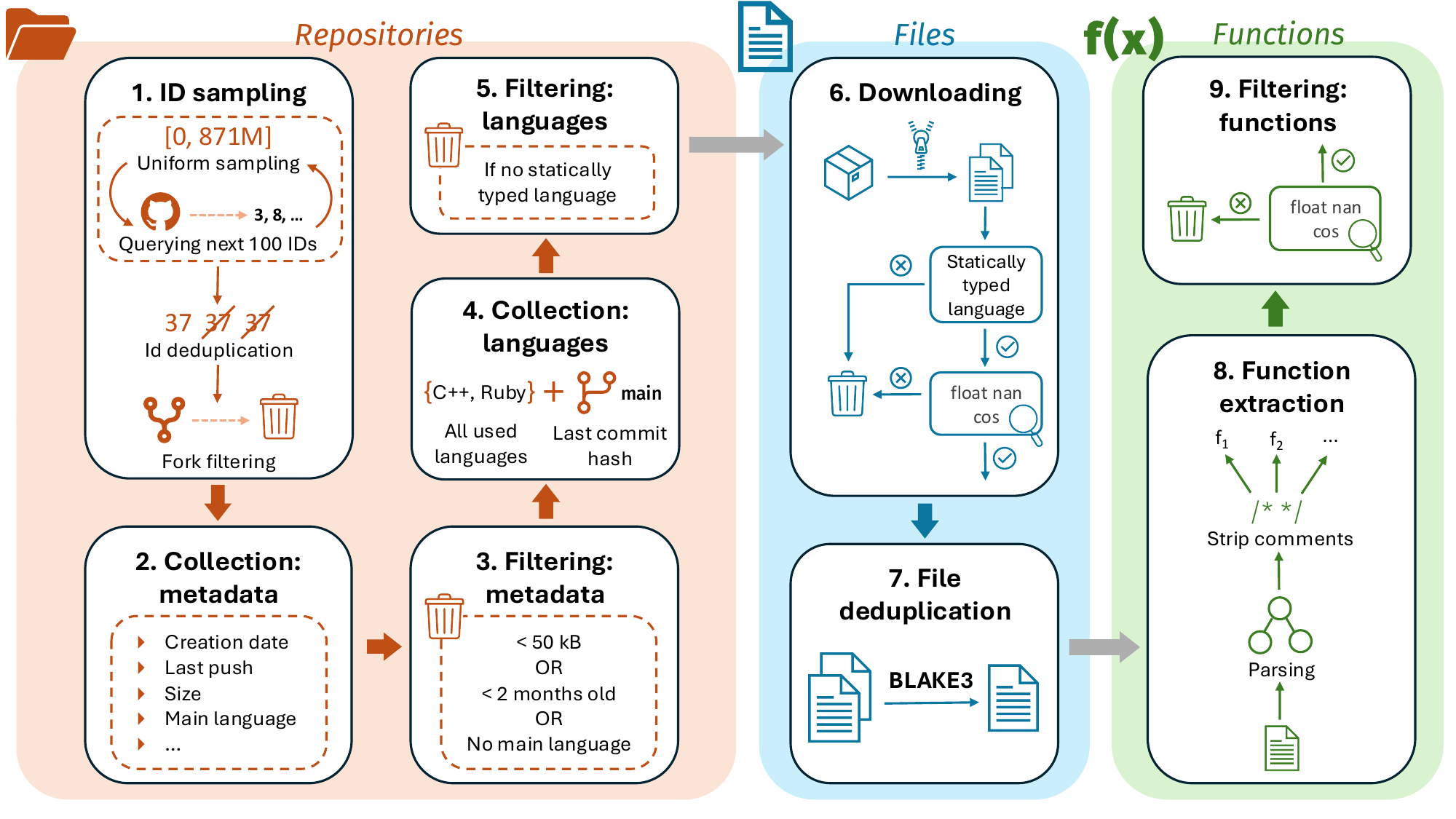}
    \caption{Multi-step methodology for identifying and analysing floating-point code}
    \Description{Repository mining pipeline: sampling, metadata filtering, language selection, downloading, deduplication, and function extraction and analysis.}
    \label{fig:pipeline}
\end{figure*}

We cannot query GitHub for non-trivial projects containing such functions
directly in part because each GitHub API call provides only limited information.
We thus design a multi-step selection process or pipeline that progressively
reduces the amount of data, increasing the level of granularity
from the repository to the file and ultimately to the function level.
An overview of the data selection pipeline is shown in \Cref{fig:pipeline}.


We begin by uniformly sampling public repositories on GitHub using GitHub's Rest
API (step~1), and by excluding forks and `trivial' projects (defined in~\Cref{sec:methodology-querying-metadata}, step~2~and~3).
For each remaining repository, we query the list of used programming languages and retrieve the
hash of the last commit on the main branch (step~4). We discard projects that do not
have any files written in statically typed languages (step~5). \emph{At the end
of this step, we have metadata about the repositories, after preliminary filtering.}

Next, we download the content of the selected repositories and inspect
individual files within these that are written in statically typed programming languages (step~6).
For these we identify floating-point usage using keywords in the source code
directly, \eg{} focusing on type annotations (\eg{} \code{double} and \code{real}) or
mathematical functions (\eg{}
\code{sin} and \code{exp}).
We discard data that we do not consider: files that are not written in
statically typed languages, files that do not contain any keyword related to
numerical computations, and files that are token-level duplicates of other files (in
any considered repository, step~7).
We then parse the remaining files, remove comments, and extract individual functions (step~8).
We retain only functions that include keywords related to floating-point computations (step~9).
In the final step we quantify usage of programming language constructs such as
loops or conditional statements fully automatically based on parse trees.


The pipeline is designed to be fully automated, reproducible, modular and extensible. The sampling
and filtering steps earlier in the pipeline are not specific to floating-point
arithmetic and can be re-used for other code studies.
We envision the floating-point specific dataset collected at the end of the
pipeline to be used, \eg{} to sample or validate benchmarks with specific
features that are used for evaluating new floating-point automated reasoning
techniques.

%% file: sections/data_collection.tex
\section{Experimental Setup}\label{sec:data-collection}

\paragraph*{Implementation}
Our approach is implemented in Rust and designed as a general-purpose framework, called Scyros, for large-scale code studies, with a focus on
reproducibility and reusability.
Data collection can be interrupted and resumed at any time without compromising reproducibility. Inputs are randomly shuffled before any pipeline step to enable subsampling. All random seeds are fixed and can be user-defined.
During language collection, the hash of the latest commit on the main branch is stored, ensuring that repository contents remain consistent across re-runs, provided it is still reachable and history was not rewritten.

Scyros can be straightforwardly adapted in a number of ways for different analyses.
For example, users can choose to analyse
local repositories instead of downloading data from GitHub. They can also
specify which file extensions or programming language to retain in the dataset,
and adapt the set of keywords for each programming language depending on the target code features of interest. Multiple
configuration files can be supplied simultaneously, enabling parallel analyses
with different settings. When downloading data, users have the option to either
delete files that do not pass filtering criteria or keep them for future use.
Each pipeline step produces a CSV file that serves as the input for the next step.
All steps can be executed independently as separate subcommands via the command-line interface. Individual steps can be skipped or performed manually by editing their corresponding CSV files without affecting other stages of the pipeline. Furthermore, additional steps can be introduced without disrupting existing functionality.
The code is fully documented, tested, and released as open-source software under the Apache 2.0 license.
We conduct the data analysis in Python.

\paragraph*{Data Collection}
We run our pipeline on an Ubuntu 24.04 machine with 128 GB RAM, 12 CPU cores at 4.95 GHz, using Rust 1.85.0 and Python 3.13.9.
We ran the different pipeline steps in an interleaved fashion. Since the exact running
time of each pipeline step depends directly on the number of sampled projects,
we report here approximate durations that are aimed to mostly illustrate the
relative differences between the pipeline steps. Steps that rely on the GitHub API
are rate-limited to 5000 requests per hour per token, hence the durations vary
depending on the number of available tokens. For our experiments, we used 8
tokens.

Sampling IDs took approximately a week, metadata collection took 2 months,
language collection took a few days, downloading the files took 3 days, removing
duplicates took 10 minutes and parsing took 60 minutes. We collected the data
between October 2024 and March 2025.

\paragraph*{Project Metadata Filtering}
We first provide statistics about steps 1--5 of our pipeline
(see~\Cref{fig:pipeline}), which are largely not specific to floating-point
arithmetic, before discussing the floating-point specific results in~\Cref{sec:results}.

After deduplication, random sampling yielded \num{125234729}
unique GitHub repository IDs, of which \Pcent{67} (\num{83615678})
were non-fork repositories (step 1). Because GitHub returns \num{100}
repository IDs per request, but metadata can only be retrieved one
repository at a time, we subsampled the IDs to make metadata
collection tractable. Metadata collection (step 2) for sampled project IDs
yielded \num{11924148} project entries.

At the time of metadata collection, \Pcent{4.6} of the projects were no longer reachable, having been deleted or made private.
Metadata for the remaining reachable projects is summarised in the first row of \Cref{tab:metadata}.
For each statistic, the table reports the minimum, median, and maximum values, as well as the first (Q1) and third (Q3) quartiles.
\begin{table}[t]
  \caption{Distribution of project metadata before and after metadata filtering, and after keyword filtering. Q1 and Q3 denote the first and third quartiles. The gray row is discussed in~\Cref{sec:results}.}
  \input{figures/metadata_table.tex}
  \label{tab:metadata}
\end{table}

In summary, the dataset is skewed towards small, recent and
inactive projects. More than \Pcent{75} of the projects are
younger than two weeks, and \Pcent{85} younger than two months.
Empty repositories account for \Pcent{19} of the dataset, while
\Pcent{49} are smaller than \SI{50}{\kilo\byte}. 
Additionally, \Pcent{12} 
of the projects are non-empty but have no primary language assigned by GitHub, 
confirming the need for filtering. A
small fraction (\Pcent{0.61}) of the projects have a negative age,
a consequence of GitHub relying on the local system time of the
machine pushing the commit, which may be incorrectly configured.


After removing projects smaller than \SI{50}{\kilo\byte}, 
that were less than \agethreshold{} months old, or contained
no source code (step 3), we were left with \num{1238741} projects
(\Pcent{11} of the initially collected metadata). We refer to
these as the ``interesting'' projects in the second row of
\Cref{tab:metadata} which shows their metadata distribution. While the
number of projects created on GitHub has grown exponentially, with
the majority in the last five years, this trend is less pronounced
in the filtered dataset. The median creation year of retained
projects is two years earlier than that of the unfiltered dataset.

The dataset is dominated by projects written primarily in non-statically typed
languages. According to GitHub's language classification, JavaScript accounts
for the largest share at
\Pcent{22}, followed by HTML (\Pcent{12}) + CSS (\Pcent{4}) and Python
(\Pcent{10}) + Jupyter Notebook (\Pcent{5}). These proportions are similar
before and after filtering, except for HTML, which drops from \Pcent{17} to
\Pcent{12}. Among statically typed languages, the most common are Java
(\Pcent{9}), \ts{} (\Pcent{7}) and C++ (\Pcent{4}). After collecting the list of
all languages used in each repository (step 4), and removing projects that do not include
any statically typed language (step 5), we are left with \num{465050} repositories,
corresponding to \Pcent{38} of the filtered dataset. This dataset forms the basis for answering our
research questions.

%% file: figures/metadata_table.tex
    \centering
    \small
    \setlength{\tabcolsep}{4pt}
    \begin{tabular}{llccccccc}
    \toprule
    & Metadata & Age & Creation & Size & \multicolumn{2}{c}{\# forks} & \multicolumn{2}{c}{\# issues} \\
    \cmidrule(lr){4-4} \cmidrule(lr){5-5} \cmidrule(lr){6-7} \cmidrule(lr){8-9}
    & & & (YY/MM) & (in kB) & > 0 & among > 0 & > 0 & among > 0 \\
    \midrule 
    \multirow{5}{*}{\rotatebox{90}{All}} &
    Min      & --19459\,\days & 70/01 & 0            & \multirow{5}{*}{\Pcent{6}}  & 1              & \multirow{5}{*}{\Pcent{6}} & 1            \\
    & Q1     & 1\,\minutes   & 20/02 & 2            &                             & 1              &                            & 1            \\
    & Median & 36\,\minutes  & 22/02 & 56           &                             & 1              &                            & 2            \\
    & Q3     & 11\,\days     & 23/08 & 928          &                             & 3              &                            & 7            \\
    & Max    & 6104\,\days   & 25/01 & 109 \millions&                             & 144 \thousands &                            & 70 \thousands\\
    \midrule  
    \multirow{5}{*}{\rotatebox{90}{Interesting}} &
    Min      & 60\,\days     & 08/02 & 50           & \multirow{5}{*}{\Pcent{20}} & 1              & \multirow{5}{*}{\Pcent{31}}& 1            \\
    & Q1     & 125\,\days    & 19/05 & 322          &                             & 1              &                            & 1            \\
    & Median & 316\,\days    & 20/10 & 1836         &                             & 2              &                            & 5            \\
    & Q3     & 925\,\days    & 22/11 & 9759         &                             & 5              &                            & 14           \\
    & Max    & 6104\,\days   & 24/10 & 105 \millions&                             & 144 \thousands &                            & 70 \thousands\\
    \midrule  
    \multirow{5}{*}{\colorbox{gray!15}{\rotatebox{90}{\parbox{1.1cm}{~With FP \newline keywords}}}} &
    Min      & 60\,\days     & 08/03 & 50           & \multirow{5}{*}{\Pcent{24}} & 1              & \multirow{5}{*}{\Pcent{30}}& 1            \\
    & Q1     & 119\,\days    & 19/01 & 372          &                             & 1              &                            & 1            \\
    & Median & 290\,\days    & 20/12 & 2102         &                             & 2              &                            & 4            \\
    & Q3     & 844\,\days    & 22/12 & 14422        &                             & 6              &                            & 13           \\
    & Max    & 6090\,\days   & 24/10 & 42 \millions &                             & 27 \thousands  &                            & 15 \thousands\\
    \bottomrule
    \end{tabular}

%% file: sections/results.tex
\section{Quantitative Analysis of Floating-Point Usage}\label{sec:results}

We address the following research questions to evaluate the results of our study and outline how we answer each of them:

\begin{itemize}[label={}, itemindent=0ex, leftmargin=0ex, itemsep=1ex]
    \item \textbf{RQ1. How prevalent is floating-point arithmetic in open-source statically typed code on GitHub?}
    The aim is to confirm (or disprove) the conventional hypothesis, thus (in)validating the motivation for much past and future research.

    \quad{} We measure prevalence at both the project and file
    levels. First, we identify how many repositories and files
    contain floating-point code. At the file level, we further
    investigate whether floating-point operations are distributed
    throughout the file or concentrated in a few functions. We
    provide the complete answer to RQ1 on page \pageref{rq1}.

    \item \textbf{RQ2. What are the characteristics of floating-point code in terms of programming language constructs and size?}
    The aim is to characterise "real-world" floating-point code to
    inform future floating-point tool development and benchmarking towards
    practical relevance.

    \quad{} We analyse the structure of floating-point functions in our dataset to
    identify commonly used programming language constructs, such as loops,
    branches, functions calls, and their combinations.

    \item \textbf{RQ3. How representative are benchmarks used in the literature to evaluate floating-point analyses of real-world floating-point code?}
    We compare functions from FPBench~\cite{damoucheStandardBenchmarkFormat2017}
    with those in our dataset based on their structural characteristics. We also
    examine how frequently functions from the GNU Scientific library that are used in the literature appear in
    real-world code.

\end{itemize}

In the following, unless otherwise specified, all reported quantities are
rounded to two significant digits to enhance readability.

\begin{table}[t]
    \caption{Proportion of projects, resp. files that contain certain types of keywords}
    \input{figures/kw_in_files_table.tex}
    \label{tab:keywords_prop}
\end{table}

\subsection{Project-Based Analysis}

At the time of downloading (step 6 in~\Cref{fig:pipeline}), \Pcent{3.8} of the
projects were unreachable, leaving \num{447209} accessible projects. From these,
we now consider only those projects that contain floating-point keywords.
Metadata for these projects is summarised in the third (grey) row
of~\Cref{tab:metadata}. While not the same, the metadata is quite similar to the
one of the ``interesting'' projects.

\Cref{tab:keywords_prop} shows the breakdown of the different floating-point
keyword categories by projects and files.
Floating-point keywords appears in \Pcent{66} of the projects. Most often these are floating-point types, which appear in \Pcent{64}
of the projects. Transcendental functions and keywords such as \code{NaN} appear
less frequently explicitly in the code, while arbitrary-precision keywords appear very rarely (\Pcent{1.1} of the projects).

For each project containing at least one floating-point keyword from a category,
we compute the proportion of statically typed files containing at least one
keyword from that category. The median and quartiles of these
proportions are shown in the second column of the table. What we observe is that
when floating-point related keywords are present in a project, they tend to be concentrated in a small
number of files.
In half of the projects that contain floating-point types, fewer than \Pcent{17}
of the files include such keywords. This proportion is even lower for
transcendental functions as well as miscellaneous and arbitrary-precision keywords.

We note that we compute these statistics before deduplication. This means that
these statistics will likely contain some library code for projects but not for
all, depending on which code has been committed to the repository.


\subsection{File-Based Analysis}
\paragraph{Presence of Keywords}
After downloading the repositories and retaining only the files written in statically typed languages with floating-point keywords, we are left with \num{13167648} files.
To estimate the false positive rate, we draw a random sample from this dataset and manually inspect each file to determine whether it contains actual floating-point code.
Using Cochran's formula \cite{cochran1977sampling} for sample size estimation in large populations, we select 385 files, ensuring statistical representativeness at a \Pcent{95} confidence level with a \Pcent{5} margin of error.
Of these 385 files, 314 contain valid floating-point code, while 71 are false positives. Among the false positives, 50 arise from keywords appearing in comments, 9 from keywords in string literals, 7 from keywords used in variable names, and 5 from other miscellaneous causes. In most cases, the keyword double appears in comments in C and C++ files.
This indicates the importance of stripping comments and ignoring string literals before performing keyword-based detection at the function level in later steps of the methodology.

Given the results of the manual inspection and using Wilson score intervals \cite{Wilson01061927}, we estimate that the proportion of true positives in our dataset lies between \Pcent{77} and \Pcent{86}, with \Pcent{95} confidence.
Assuming that the presence of false positives in files is independent across files, the probability that project~$i$ contains at least one file with actual floating-point code is
\[ P_i = 1 - (1 - p)^{k_i} \]
where $p$ is the estimated true positive rate, and $k_i$ is the number of files with floating-point keywords in project~$i$.
Note that the independence assumption does not fully reflect reality, as files within the same project are not independent.
Taking the expectation of $P_i$ over all projects, we estimate that between \Pcent{62} and \Pcent{64} of the projects in our dataset contain floating-point code, with \Pcent{95} confidence.

That said, some files may have been missed due to the use of type-aliases or user-defined wrappers, so the numbers may be even higher.
To assess the extent of this issue, we randomly sample 385 projects and manually inspect whether they redefine floating-point types.
Among these, 22 projects do redefine floating-point types. Example of such aliases include declarations like \code{type money = number} in \ts{},
or conditional typedefs that map \code{float} and \code{double} to a common type \code{Float} to select precision at compile time with macros in C++.
In all observed cases, these aliases are used alongside standard floating-point types rather than replacing them entirely, making the precise impact of this phenomenon difficult to quantify. 
An additional 10 projects define wrappers around floating-point values, typically simple structural wrappers in abstract syntax tree nodes.
Over the full dataset, we therefore estimate with \Pcent{95} confidence that between \Pcent{5.9} and \Pcent{11.5} of projects are affected by floating-point type redefinitions or wrappers.

\paragraph{De-Duplication}
A substantial fraction of the dataset is redundant, 
with \Pcent{63} of files identified as token-level duplicates and \Pcent{61} as exact duplicates.
\ts{} exhibits the highest duplication rate, with declaration files duplicated on average 17 times and source files twice.
C follows, with 3.4 copies per source file and 2.8 per header file. The most
duplicated file originates from Angular, a \ts{} web application framework, and appears \num{27136} times (\Pcent{0.70} of \ts{} files in the dataset).
Of the 100 most duplicated files, 98 are written in \ts{}. Collectively, these files account for \Pcent{9.8} of 3.9 million \ts{} files in the dataset.
The most duplicated C files are mainly hardware drivers implementations.

\paragraph{Most Common Programming Languages}
For parsing, we focus on the most commonly used languages in the dataset. 
However, using a single metric, such at the number of lines of code or files, to compare different programming languages is misleading.
Some languages are inherently more verbose, while others require code to be split across multiple files.
In addition, the rate of false positives varies by language. 
Therefore, we focus on the orders of magnitude, and avoid drawing conclusions based on exact numbers or ordering between languages with similar proportions.

To better understand each language's contribution and mitigate potential biases, we separate C and C ++ header files from C source files.
We distinguish between C headers (\code{.h} and \code{.inc}), which can also be used in C++, and C++-specific headers (\code{.hpp, .hh, .hxx, .h++, .ipp, .tpp, .tcc}).
Similarly, we treat \ts{} declaration files (\code{.d.ts}) separately from \ts{} source files (\code{.ts}).

We consider several metrics to identify which languages are most used for floating-point computations in our dataset. This comparison is illustrated in \Cref{fig:languages}. Each bar represents the proportion of files or words in the dataset that contain floating-point keywords, grouped by language. The first and third bars include all categories of keywords, while the others exclude keywords corresponding to floating-point types. We detail the choice of metrics and the corresponding results below.

\begin{figure*}[t]
    \centering
    \begin{adjustbox}{width=1\textwidth}
    \input{figures/languages.tex}
    \end{adjustbox}
    \caption{Proportion of files and words in the dataset with floating-point keywords for the most used languages}
    \Description{Stacked bars showing proportions of C, C++, Java, TypeScript, C#, and Go in files and words containing floating-point keywords. All bars go beyond 92\% cumulatively, with C and C++ dominating.}
    \label{fig:languages}
\end{figure*}

\begin{itemize}[label={}, itemindent=0ex, leftmargin=0ex, itemsep=1ex]
    \item \textbf{Proportion of files with floating-point keywords (first bar):}
    Counting the number of files, most of the code with any floating-point
    keyword is written in Java (\Pcent{20}), while C++-specific headers are less
    common (\Pcent{2.2}).
    The high number of Java files can be partly explained by the convention of isolating each class in its own file.
    In \ts{}, the use of the \texttt{number} type to represent both integers and floating-point values inflates the keyword count.
    To mitigate this effect, we also report the proportion of files excluding floating-point types and including only transcendental functions, arbitrary-precision and miscellaneous keywords.

    \item \textbf{Proportion of files with transcendental functions, arbitrary precision and
    miscellaneous keywords (second bar):} In this case, C++ source files make up
    most of the code (\Pcent{28}). Notably, the proportion of \ts{} drops to
    \Pcent{4.6} for source files and \Pcent{0.62} for declaration files.

    \item \textbf{Proportion of words of files with floating-point keywords (third bar):}
    When counting words instead of files, the proportion of Java code drops as expected (\Pcent{9.5}),
    and C and C++ source code has the largest share (\Pcent{28} and \Pcent{23}, resp.).

\end{itemize}

From this observation we conclude that C, C++, Java, \ts{}, C\# and Go are the most commonly used languages for floating-point computations in our dataset. 
We do \textit{not} interpret this as evidence that these languages are inherently better suited for writing floating-point code, nor that, relative to their general usage, they are more frequently employed for writing numerical code than other languages.
Regardless of the metric used, these six languages collectively account for more than \Pcent{92} of the dataset. We also explored other metrics, such as the number of lines of code and the exclusion of specific categories of keywords, but these variations did not lead to any substantial change in the overall trends.

\subsection{Per-Function Analysis}

Based on the observation of the previous section, we parse source files written
in C, C++, Java, \ts{}, C\# and Go and extract functions containing
floating-point keywords (steps 8 and 9 in~\Cref{fig:pipeline}). 
We again estimate the false-positive rate by manually inspecting a representative sample of 385 extracted functions. 
Among these, 368 functions (\Pcent{96}) contain valid floating-point code. 
The remaining false positives fall into two categories: nine cases in which floating-point keywords are used as variable or library names, and eight functions that include floating-point types only in parsing-related code. 
Based on this sample, we estimate with \Pcent{95} confidence that the true positive rate of our extraction lies between \Pcent{93} and \Pcent{98}.

For each function, we count the number of lines of code, words, parameters,
floating-point keywords of the different categories, loops, conditional statements,
function calls and maximum nesting depth of these constructs.
This is the main dataset we publicly release. Depending on the features a new verification tool is targeting, the developer can filter the dataset to sample relevant functions leveraging the above metrics.

\begin{table}[t]
\caption{Statistics of extracted functions}
\input{figures/extracted_fn.tex}
\label{tab:functionsperfile}
\end{table}

\Cref{tab:functionsperfile} shows the number of files per language from which we extract functions (first column), and how many of these files contain at least one function (second column). 
The third column reports the proportion of those files that include at least one function with a floating-point keyword.
The last column gives the average number of such functions per file (among files with at least one), along with the standard deviation.
We observe that most languages have a high proportion of files with at least one function, except \ts{} (\Pcent{58}), which includes many small files that export only constants.
As previously described, most false positives in our manual inspection of files come from keywords in comments or string literals, but these are no longer present in our function dataset.
The proportion of files with at least one function using a floating-point keyword aligns with the number of true positives we found manually during our per-file inspection.
Due to the high variation in the data, we cannot draw a clear conclusion about the proportion of floating-point functions per file.

\begin{rqanswer}\label{rq1}
    \textbf{Answer to RQ1:} Floating-point code is prevalent in open-source code.
    With \Pcent{95} confidence, we estimate that more than \Pcent{62} of projects contain floating-point code.
    Within projects, this code is often concentrated in a small number of files.
    We observe no clear trend in the proportion of functions using floating-point code within individual files.
\end{rqanswer}

\paragraph{Structure of Floating-Point Functions}
Our final dataset consists of \num{9945253} functions, of which \num{1168217}
are written in C (\Pcent{12}), \num{4176635} in C++ (\Pcent{42}), \num{2574341}
in Java (\Pcent{26}), \num{723609} in \ts{} (\Pcent{7}),
\num{1009329} in C\# (\Pcent{10}), and \num{293122} in Go (\Pcent{3.0}).
\Cref{tab:functions} reports on function complexity and the use of different
language constructs in this dataset. For each construct, we report the
proportion of functions that use it, along with the median and interquartile
range (IQR, \ie{} the difference between the 75th and 25th percentiles)
of its occurrences among those functions. For each keyword category, we report
the proportion of functions using at least one keyword from that category.
For illustration purposes, \Cref{fig:misc-kw} shows a randomly selected
function with a miscellaneous keyword (\code{isFinite}); note that no other floating-point
related keywords appear.

We note that the number of floating-point parameters and sizes of functions is
relatively small, across all languages. C code tends to have the `longest'
functions. For each language, more functions contain function calls than conditional
statements, and fewer contain loops.{}
\begin{table*}[t]
    \centering
    \small
    \caption{Function statistics per function. FPB = FPBench. Med denotes median and IQR the interquartile range, computed out of positive occurrences only.}

    \begin{subtable}{1\textwidth}
        \input{figures/functions_fp.tex}
        \subcaption{Floating-point code}
        \label{tab:functions}
    \end{subtable}

    \begin{subtable}{1\textwidth}
        \input{figures/functions_all.tex}
        \subcaption{All code}
        \label{tab:non_num_functions}
    \end{subtable}

\end{table*}

\begin{figure*}[t]
    \centering
\begin{lstlisting}[style=highlight,language=Go]
func writeSVG(writer io.Writer, f biFunc) {
	fmt.Fprintf(writer, "<svg xmlns='http://www.w3.org/2000/svg' " +
		"style='stroke: grey; fill: white; stroke-width=0.7' " +
		"width='%d' height='%d'>\n", width, height)

	for i := 0; i < cells; i++ {
		for j := 0; j< cells; j++ {
			ax, ay := corner(i + 1, j, f); bx, by := corner(i, j, f)
			cx, cy := corner(i, j + 1, f); dx, dy := corner(i + 1, j + 1, f)
			if	!isFinite(ax) || !isFinite(ay) || !isFinite(bx) || !isFinite(by) ||
				  !isFinite(cx) || !isFinite(cy) || !isFinite(dx) || !isFinite(dy) {
				continue
			}
			r, g, b := color(i, j, f)
			fmt.Fprintf(writer, "<polygon points = '%g,%g,%g,%g,%g,%g,%g,%g'
			 stroke = 'rgb(%d, %d, %d)'/>\n", ax, ay, bx, by, cx, cy, dx, dy, r, g, b)
		}
	}
}
\end{lstlisting}
    \caption{ Randomly picked Go floating-point function in our dataset that contains miscellaneous keywords \cite{writeSVG}.}
    \label{fig:misc-kw}
    \Description{Go function example showing use of isFinite to filter non-finite floating-point values during SVG generation.}
\end{figure*}
\begin{figure*}[t]
    \centering
\begin{lstlisting}[style=highlight,language=C++]
void RolloutStorage::compute_returns(torch::Tensor next, bool use_gae, float gamma, float tau) {
    if (use_gae) {
        value_predictions[-1] = next;
        torch::Tensor gae = torch::zeros({rewards.size(1), 1}, torch::TensorOptions(device));
        for (int step = rewards.size(0) - 1; step >= 0; --step) {
            auto delta = rewards[step] + gamma * value_predictions[step + 1] * masks[step + 1] - value_predictions[step];
            gae = delta + gamma * tau * masks[step + 1] * gae;
            returns[step] = gae + value_predictions[step];
        }
    }
    else {
        returns[-1] = next;
        for (int step = rewards.size(0) - 1; step >= 0; --step)
            returns[step] = returns[step + 1] * gamma * masks[step + 1] + rewards[step];
    }
}
\end{lstlisting}
    \caption{ Randomly picked C++ floating-point function in our dataset that contains both a loop and a conditional statement \cite{compute_returns}.}
    \label{fig:ex-if-while}
    \Description{C++ function computing intermediate floating-point points along a line segment and returning them as cell coordinates.}
\end{figure*}

In addition to the information in the table, we also analyse combinations of
constructs: in C\#, \Pcent{10} of functions include loops,
conditionals, and function calls within a single function, compared to
\Pcent{26} in C. When a function contains a loop, a conditional is also present
in \Pcent{69} (Java) to \Pcent{87} (Go) of functions. \Cref{fig:ex-if-while} is an example of a randomly selected C++ function from our dataset that contains both a loop and a conditional statement.

We compare these results with those obtained from non-numerical code.
We uniformly sample \num{10000} projects from our dataset and apply the same methodology, yielding \num{9708183} functions.
Characteristics of these functions are summarised in \Cref{tab:non_num_functions}.

Compared to functions sampled from all projects, floating-point functions tend to be larger and contain loops more frequently.
The larger average size of numerical functions may partly result from our keyword-based filtering, as larger functions are more likely to contain such keywords.
The increased presence of loops could be a consequence of the larger average size of numerical functions.
To assess whether function size alone accounts for the observed disparity in loop presence, we perform a controlled analysis.
We fit a binary logistic regression model~\cite{hosmerAppliedLogisticRegression2013} predicting the presence of a loop as a function of the logarithm of function size and whether the function contains floating-points.
After controlling for size, numerical functions remain significantly more likely to contain loops, with coefficients ranging from 0.1215 $\pm 0.003$ for C++ to 0.6395 $\pm 0.007$ for C\#, all with $p < 0.001$.
According to the model, for functions of equal size, the odds of containing a loop are on average \Pcent{13} (C++) to \Pcent{90} (C\#) higher in floating-point functions than in the general population.
Although these results should be interpreted cautiously, they indicate that floating-point functions exhibit a systematically higher prevalence of loops beyond what can be explained by function size alone.
The models achieve ROC AUC values between 0.79 (C) and 0.86 (Java). We do not observe a comparable effect for the presence of conditionals.

\paragraph{Mixed-Precision Code} 
We also analyse the use of different floating-point precisions within functions in our dataset.
\Cref{tab:mixed_precision} reports, for each precision, the proportion of functions in which a corresponding type keyword appears, 
as well as the proportion of functions that use keywords from multiple precisions within the same function. 
We omit \ts{} from this analysis, as it provides only a single floating-point type (\code{number}).
For C and C++, we additionally report half precision (16-bit) and quad precision (128-bit), which are supported by these languages.

Single (32-bit) and double (64-bit) precision are the most commonly used across all languages, athough their relative prevalence varies substantially by language.
This variation may be partly explained by differences in the libraries commonly used in each ecosystem.
For example, in C\#, where \Pcent{65} of functions use single precision, 
the Unity library predominantly relies on 32-bit floating-point data types such as \code{Vector3} and \code{Quaternion}. 
In contrast, Go uses double-precision keywords in \Pcent{82} of functions,
 consistent with common practice in Go ecosystems where numerical libraries often operate on 64-bit floating-point values. 
Half (16-bit) and quad (128-bit) precision, as well as mixed-precision code, occur only rarely.

\begin{table}[tb]
    \caption{Distribution of floating-point precision keywords used in functions per language: half (16-bit), single (32-bit), double (64-bit), quad (128-bit).}
    \input{figures/mixed_prec.tex}
    \label{tab:mixed_precision}
\end{table}

\paragraph{Parsing-Errors}
Because not every source file is successfully parsed by Tree-sitter,
we investigate how many files are affected by parsing errors and whether this biases our dataset toward parser-supported constructs.
Tree-sitter recover from parsing errors by skipping invalid tokens and inserting placeholder nodes where syntactic elements are expected.
Across the full dataset, Tree-sitter generate such error-recovery nodes in 
\Pcent{16} of the files. The languages most affected are C and C++, with \Pcent{53} and \Pcent{33} of files affected, respectively. 
Java and Go are significantly less affected, with \Pcent{0.50} and \Pcent{0.23} of files, respectively.

To assess the impact of these errors and characterise their causes,
 we manually inspect a sample of 97 files affected by parsing errors (corresponding to a Cochran sample size with a \Pcent{10} margin of error). 
 In 90 out of 97 cases, parsing errors do not affect function extraction, \ie{} no functions are skipped.
Based on this observation, we estimate with \Pcent{95} confidence that between \Pcent{0.57} and \Pcent{2.3} of files in the full dataset have functions missed due to parsing errors.
In most cases (74 out of 97), parsing errors stem from the use of preprocessor directives in C and C++ code. 
A common scenario involves conditional compilation directives (\eg{} \code{#ifdef}), which may introduce unmatched or duplicated closing braces at the end of a function.
Because preprocessor directives can only be resolved with full knowledge of the surrounding code, 
they are difficult to be handled reliably when analyzing files in isolation.
Other sources of parsing errors include macros used as keywords (\eg{} replacing \code{static} or \code{inline}) or macros that expand to language constructs such as \code{for} loops. 
We also observe cases in which Tree-sitter cannot correctly parse macro arguments without resolving macro expansion, for example when those arguments contain statements or type declarations. 
Finally, some errors (11 out of 97) arise from syntactic forms not supported by the Tree-sitter grammar, such as \code{new ptr*[n]}.

\begin{rqanswer}\label{rq2}
    \textbf{Answer to RQ2:} In our dataset, conditional statements appear in many functions, up to \Pcent{63} in Go.
    This proportion increases when loops are present, although loops themselves are less frequent.
    For functions of equal size, floating-point functions are more likely to contain loops than functions sampled from all code.
    Function calls are also common in floating-point code, and typically these are not calls to transcendental functions.
    Special values like NaN or infinity appear rarely but are present.
    Despite these complexities, functions tend to be small, highlighting the
    importance of modularity in floating-point reasoning techniques. 
    Single- and double-precision floating-point types are the most commonly used across all languages, whereas other precisions,
    arbitrary precision, and mixed-precision code are relatively rare.
\end{rqanswer}

\paragraph{Existing Benchmarks}
We compare our dataset to FPBench~\cite{damoucheStandardBenchmarkFormat2017}, a
benchmark suite of 131 floating-point functions written in Racket and designed
for verification tools. FPBench functions can be exported to several supported
languages. We export them to C and run our parser to extract statistics (see
last row of \Cref{tab:functions}).

The FPBench functions have notably different characteristics: conditional
statements and function calls appear less frequently, while calls to
transcendental functions appear more often. The combination of a loop,
conditional statement and a function call occurs only \Pcent{1.5} of the time
(compared to \Pcent{26} in C). When a function contains a loop, a conditional is
also present in \Pcent{27} of the cases (compared to over \Pcent{69} in
real-world code).
FPBench does align with our dataset in terms of the number of function parameters. The
word count may be biased by the code being automatically exported to C from Racket
and thus not written like a human would, \eg{} with fewer intermediate variables.

We presume that this difference arises because the FPBench benchmarks
    are drawn from existing literature and consist of hand-picked programs that
    current tools can already analyse successfully. In particular, many static
    floating-point reasoning tools offer limited support for conditionals,
    loops, and arbitrary function calls, which is reflected in the low
    prevalence of such constructs in FPBench.


The GNU Scientific Library (GSL) is also frequently used in the literature for benchmarking purposes.
Guo et al. \cite{guoEfficientGenerationErrorinducing2020} use a subset of nine functions from the GSL Stats module. Other studies employ various subset of the GSL Special Functions module, ranging from 14 selected functions to the entire module \cite{barrAutomaticDetectionFloatingpoint2013,zouGeneticAlgorithmDetecting2015, yiEfficientAutomatedRepair2019,zouDetectingFloatingpointErrors2019,miaoCompilerNumericalInconsistencies2024}.

We assess the prevalence of GSL usage in our dataset by computing the proportion
of files and functions that use functions from the SF (special
functions), Stats and Math (common mathematical functions) modules. We perform
this analysis by supplying additional JSON files to the downloader and parser,
following the same approach used for identifying floating-point keywords. For
each module, we list all available functions and check for their presence in the
dataset. This process also shows the flexibility of our implementation 
in supporting additional analyses on an existing dataset.

The results are summarised in \Cref{tab:gsl}. We observe that GSL modules
are not widely used in the dataset. Only \Pcent{0.084} of C files and
\Pcent{0.023} of C++ files with floating-point keywords include functions from
the GSL Math module. The Stats and SF modules appear even less frequently. In
comparison, the math.h library is included in \Pcent{17} of C and \Pcent{14} of C++ files with
floating-point keywords. Usage at the function level is similarly sparse as
only \Pcent{0.13} of C floating-point functions include calls to GSL SF
functions.

\begin{table}[tb]
    \caption{Files and functions using GSL functions}
 \input{figures/gsl.tex}
\label{tab:gsl}
\end{table}

\begin{rqanswer}\label{rq3}
    \textbf{Answer to RQ3:} FPBench functions are representative of real-world floating-point code in terms of the number of parameters but not in terms of complexity. 
    Conditionals are rare, and function calls, when present, typically involve transcendental functions that are over-represented.
    Functions from the GNU Scientific Library that have been used for evaluation appear rarely in our dataset.
\end{rqanswer}



\subsection{Threats to Validity}
A first threat to validity concerns the generalisability of our findings.
By sampling data from GitHub and carefully filtering it, we analyse real-world code.
However, industrial, closed-source software or open-source software from other sources than GitHub may exhibit characteristics different from those observed in open-source projects on GitHub.
Large parts of our methodology and implementation are reusable on other data sources or within industrial case studies, which could help assess the extent of these differences in future work.

Another potential threat relates to the accuracy of metadata provided by GitHub.
Commit timestamps depend on the local system time of the machine pushing the
commit, which may be misconfigured. As a result, the recorded creation and last
push dates of a project may not be reliable, which could affect our filter.
A partial mitigation is to use the timestamp of the first commit instead of the
reported project creation date, since an incorrect system clock would likely produce a consistent offset. 
However, this information is not included in the metadata returned by the API, 
and retrieving it explicitly would require an additional request per repository, 
effectively doubling the number of API calls.
Similarly, GitHub's language classification is based on heuristics and may
misidentify the main language of a project. This issue also applies to our own
language detection, which relies on file extensions. Since some extensions are
shared between multiple languages, misclassification is possible. Our manual
inspection suggests, however, that this affects only a small fraction of the
dataset.

A third threat relates to the keyword-based approach we use to detect floating-point code.
Some transcendental functions appear as calls to external libraries, which we do not capture, potentially skewing our statistics. 
We also do not identify operations on complex numbers, which often rely on floating-point arithmetic.
Abstractions such as type aliases, structures, and classes may also complicate detection of floating-point types, resulting in false negatives.
We mitigate this by including a broader set of keywords, such as transcendental functions and special floating-point values, although this only partially addresses the issue.
Keyword-based heuristics may also produce false positives. We reduce this risk by removing comments and ignoring string literals, as confirmed through manual inspection.

Duplicate removal represents another potential threat. Our approach eliminates token-level duplicates from the dataset, but does not remove near-duplicates, \ie{} files that differ only by a few characters or lines.
Such near-duplicates can arise from different versions of the same library.
Prior work by Lopes et al. \cite{lopesDejaVuMapCode2017} addresses this issue using token-similarity thresholds.
However, it remains unclear how to define an appropriate threshold or how to assess its validity in our setting. 
We leave the investigation of near-duplicate detection for future work.

Finally, the tooling itself may contain bugs, which could introduce bias. We mitigate this risk through unit and integration testing. Additionally, when we perform uniform sampling, we manually verify that the resulting distribution is consistent with expectations.

%% file: figures/kw_in_files_table.tex
\centering
\small  

\begin{tabular}{@{}lcccc@{}}
\toprule
\multirow{2}{*}{\textbf{Keyword category}} & \multirow{2}{*}{\textbf{Projects}} &  \multicolumn{3}{c}{\textbf{Files}} \\
\cmidrule(lr){3-5}
& & Q1 & Median & Q3 \\
 \midrule
Primitive floating-point types     & \Pcent{64} & \Pcent{7.1} & \Pcent{17}  & \Pcent{35}  \\
Transcendental functions & \Pcent{20} & \Pcent{1.3} & \Pcent{3.5} & \Pcent{9.5} \\
Miscellaneous keywords           & \Pcent{21} & \Pcent{1.4} & \Pcent{3.1} & \Pcent{7.1} \\
Arbitrary-precision keywords      & \Pcent{1.1} & \Pcent{0.13} & \Pcent{1.4} & \Pcent{3.6} \\
\midrule
\textbf{Any} & \textbf{\Pcent{66}} & \textbf{\Pcent{7.5}} & \textbf{\Pcent{18}} & \textbf{\Pcent{36}}\\
\bottomrule
\end{tabular}

%% file: figures/languages.tex
\pgfplotstableread{ 
Label                                             C    CH   Cpp  CppH Java TS   TSDecl CS   Go
{\small Words (excluding type keywords)} 27.8 11.9 25.1 1.9  7.3  4.0  3.3    3.8  8.4
{\small Words (all keywords)}                     22.9 13.3 27.7 1.8  9.5  5.6  3.9    4.9  5.3
{\small Files (excluding type keywords)} 16.3 10.9 27.7 2.6  18.4 4.6  0.6    6.5  4.6
{\small Files (all keywords)}                     10.3 13.3 17.8 2.2  20.2 14   3.0    10.6 2.7
    }\testdata

    \begin{tikzpicture}

    \begin{axis}[
    xbar stacked,   
    xmin=0,         
    xmax=100,      
    ytick=data,     
    legend style={at={(0.35,-0.3)},
      anchor=north,legend columns=-1},
    yticklabels from table={\testdata}{Label},  
    bar width=12pt,
    width=13cm,
    y=28.5pt,
    yticklabel style={
            text width=3.2cm,
        },
    xlabel={\small Proportion of code (\%) in},
    ]
    \addplot [fill=cbred] table [x=C, meta=Label,y expr=\coordindex] {\testdata};   
    \addplot [fill=cbred,postaction={pattern=north east lines}] table [x=CH, meta=Label,y expr=\coordindex] {\testdata};
    \addplot [fill=cborange] table [x=Cpp, meta=Label,y expr=\coordindex] {\testdata};
    \addplot [fill=cborange,postaction={pattern=north east lines}] table [x=CppH, meta=Label,y expr=\coordindex] {\testdata};
    \addplot [fill=cbyellow] table [x=Java, meta=Label,y expr=\coordindex] {\testdata};
    \addplot [fill=cbgreen] table [x=TS, meta=Label,y expr=\coordindex] {\testdata};
    \addplot [fill=cbgreen,postaction={pattern=north east lines}] table [x=TSDecl, meta=Label,y expr=\coordindex] {\testdata};
    \addplot [fill=cbcyan] table [x=CS, meta=Label,y expr=\coordindex] {\testdata};
    \addplot [fill=cbblue] table [x=Go, meta=Label,y expr=\coordindex] {\testdata};
    \legend{C, C Headers, C++, C++ Headers, Java, TypeScript, TypeScript Headers, C\#, Go}

    \end{axis}
    \end{tikzpicture}

%% file: figures/extracted_fn.tex
\begin{threeparttable}[t]
    \centering
    \small 
    
    \setlength{\tabcolsep}{10pt}
\begin{tabular}{@{}lcccc@{}}
  \toprule
\textbf{Lang} & \textbf{\# Files}\tnote{a} & \textbf{Files w/ fn}\tnote{b} & \textbf{Files w/ fn w/ kw}\tnote{c} & \textbf{\# Fn w/ kw}\tnote{d} \\
 \midrule
C             & \ku{508}          & \Pcent{97}           & \Pcent{62}                 & \Pcent{53} $\pm$ \Pcent{40}       \\
C++           & \ku{878}          & \Pcent{99}           & \Pcent{81}                 & \Pcent{41} $\pm$ \Pcent{33}       \\
C\#           & \ku{521}          & \Pcent{89}           & \Pcent{73}                 & \Pcent{41} $\pm$ \Pcent{32}       \\
Go            & \ku{134}          & \Pcent{95}           & \Pcent{70}                 & \Pcent{31} $\pm$ \Pcent{31}       \\
Java          & \ku{995}         & \Pcent{93}           & \Pcent{82}                 & \Pcent{43} $\pm$ \Pcent{34}       \\
TS            & \ku{690}          & \Pcent{58}           & \Pcent{71}                 & \Pcent{48} $\pm$ \Pcent{33}       \\
\bottomrule
\end{tabular}
\begin{tablenotes}
\small
\item [a] Number of files
\item [b] Proportion of files that have at least one function
\item [c] Proportion of files that have at least one function with a floating-point keyword (out of those with at least a function)
\item [d] Average proportion of functions with floating-point keywords per file (out of those with at least such a function), and standard deviation
\end{tablenotes}
\end{threeparttable}

%% file: figures/functions_fp.tex
        \begin{adjustbox}{width=1\textwidth}
            \begin{tabular}{@{}lc@{\hspace*{1ex}}cc@{\hspace*{1ex}}cc@{\hspace*{1ex}}c@{\hspace*{1ex}}cc@{\hspace*{1ex}}c@{\hspace*{1ex}}cc@{\hspace*{1ex}}c@{\hspace*{1ex}}cc@{\hspace*{1ex}}c@{\hspace*{1ex}}c@{\hspace*{1ex}}c@{}}
            \toprule
            \multirow{2}{*}{\textbf{Lang}} &
            \multicolumn{2}{c}{\textbf{Words}} &
            \multicolumn{2}{c}{\textbf{Params}} &
            \multicolumn{3}{c}{\textbf{Loops}} &
            \multicolumn{3}{c}{\textbf{Conditionals}} &
            \multicolumn{3}{c}{\textbf{Function calls}} &
            \multicolumn{4}{c}{\textbf{Keywords}} \\
            \cmidrule(lr){2-3} \cmidrule(lr){4-5} \cmidrule(lr){6-8} \cmidrule(lr){9-11} \cmidrule(lr){12-14}  \cmidrule(lr){15-18}
                    &  med  & IQR & med & IQR & > 0   & med & IQR & > 0 & med & IQR & > 0 & med & IQR & Types & Trasc & Misc & MPFR \\
            \midrule
            C       & 61 & 121  & 3 & 4 & \Pcent{36} & 2 & 2 & \Pcent{57} & 3 & 6 & \Pcent{84} & 5 & 12 & \Pcent{91} & \Pcent{10} & \Pcent{7.1} & \Pcent{0.38}\\
            C++     & 42 & 94  & 2 & 2 & \Pcent{22} & 2 & 2 & \Pcent{44} & 3 & 5 & \Pcent{76} & 6 & 14 & \Pcent{95} & \Pcent{6} & \Pcent{5.8} & \Pcent{0.06}\\
            C\#     & 34 & 63  & 2 & 2 & \Pcent{15} & 1 & 1 & \Pcent{42} & 2 & 4 & \Pcent{73} & 3 & 7  & \Pcent{97} & \Pcent{5.7} & \Pcent{3.5} & \Pcent{0.01}\\
            Go      & 42 & 97  & 2 & 2 & \Pcent{29} & 2 & 2 & \Pcent{63} & 2 & 4 & \Pcent{85} & 6 & 14 & \Pcent{97} & \Pcent{3.4} & \Pcent{5.2} & \Pcent{0.48}\\
            Java    & 28 & 61  & 1 & 2 & \Pcent{20} & 1 & 1 & \Pcent{37} & 2 & 3 & \Pcent{73} & 5 & 10 & \Pcent{98} & \Pcent{4.6} & \Pcent{3.9} & \Pcent{0.002}\\
            TS      & 27 & 43  & 1 & 1 & \Pcent{15} & 1 & 1 & \Pcent{44} & 2 & 3 & \Pcent{83} & 3 & 6  & \Pcent{98} & \Pcent{1.6} & \Pcent{4.0} & ---\\
            \midrule
            FPB     & 23 & 26  & 2 & 2 & \Pcent{12} & 1 & 0 & \Pcent{6.9} & 1 & 1 & \Pcent{61} & 2 & 1  & \Pcent{100} & \Pcent{38} & \Pcent{0} & \Pcent{0}\\
            \bottomrule
            \end{tabular}
            \end{adjustbox}

%% file: figures/functions_all.tex
\centering
        \begin{tabular}{@{}lccccccccccccc@{}}
        \toprule
        \multirow{2}{*}{\textbf{Lang}} &
        \multicolumn{2}{c}{\textbf{Words}} &
        \multicolumn{2}{c}{\textbf{Params}} &
        \multicolumn{3}{c}{\textbf{Loops}} &
        \multicolumn{3}{c}{\textbf{Conditionals}} &
        \multicolumn{3}{c}{\textbf{Function calls}}\\
        \cmidrule(lr){2-3} \cmidrule(lr){4-5} \cmidrule(lr){6-8} \cmidrule(lr){9-11} \cmidrule(lr){12-14}
                &  med  & IQR & med & IQR & > 0   & med & IQR & > 0 & med & IQR & > 0 & med & IQR \\
        \midrule
        C       & 42 & 60  & 2 & 2 & \Pcent{18} & 1 & 1 & \Pcent{66} & 2 & 3 & \Pcent{91} & 4 & 6 \\
        C++     & 26 & 40  & 2 & 2 & \Pcent{13} & 1 & 1 & \Pcent{37} & 2 & 3 & \Pcent{81} & 3 & 7 \\
        C\#     & 23 & 27  & 1 & 2 & \Pcent{5.5} & 1 & 1 & \Pcent{29} & 2 & 2 & \Pcent{74} & 2 & 3  \\
        Go      & 28 & 45  & 2 & 2 & \Pcent{15} & 1 & 1 & \Pcent{54} & 2 & 2 & \Pcent{80} & 4 & 8 \\
        Java    & 13 & 25  & 1 & 1 & \Pcent{9.9} & 1 & 1 & \Pcent{24} & 1 & 2 & \Pcent{61} & 3 & 6 \\
        TS      & 17 & 28  & 1 & 2 & \Pcent{6.8} & 1 & 0 & \Pcent{33} & 1 & 2 & \Pcent{76} & 2 & 4  \\
        \bottomrule
        \end{tabular}

%% file: figures/mixed_prec.tex
\begin{threeparttable}[t]
    \centering
    \small 

    \setlength{\tabcolsep}{8.5pt}
    \begin{tabular}{@{}lccccc@{}}
    \toprule
    \multirow{2}{*}{\textbf{Module}} & \multicolumn{4}{c}{\textbf{Precision}} & \multirow{2}{*}{\textbf{Mixed-Precision}}  \\
    \cmidrule(lr){2-5}
                     & Half  & Single & Double & Quad &          \\
    \midrule
    C                & \Pcent{2.7}                   & \Pcent{41}    & \Pcent{52} & \Pcent{1.9} & \Pcent{6.5}  \\
    C ++             & \Pcent{0.1}                   & \Pcent{60}    & \Pcent{38} & \Pcent{0.7} & \Pcent{3.9}    \\
    C\#              & ---                         & \Pcent{65}    & \Pcent{35} & ---  & \Pcent{2.7}      \\
    Go               & ---                         & \Pcent{23}    & \Pcent{82} & ---  & \Pcent{8.5}      \\
    Java             & ---                         & \Pcent{37}    & \Pcent{65} & ---  & \Pcent{3.8}      \\
    \bottomrule
    \end{tabular}
    \end{threeparttable}

%% file: figures/gsl.tex
    \begin{threeparttable}[t]
    \centering
    \small 
    \setlength{\tabcolsep}{20pt}
    \begin{tabular}{@{}lcccc@{}}
\toprule
\multirow{2}{*}{\textbf{Module}} & \multicolumn{2}{c}{\textbf{Files}} & \multicolumn{2}{c}{\textbf{Functions}}           \\
\cmidrule(lr){2-3}\cmidrule(lr){4-5}
                                 & C                                  & C++            & C             & C++             \\
 \midrule
GSL Math\tnote{a}                & \Pcent{0.084}                      & \Pcent{0.023}  & \Pcent{0.063} & \Pcent{0.0073}  \\
GSL SF\tnote{b}                  & \Pcent{0.071}                      & \Pcent{0.027}  & \Pcent{0.11}  & \Pcent{0.010}    \\
GSL Stats\tnote{c}               & \Pcent{0.022}                      & \Pcent{0.0034} & \Pcent{0.013} & \Pcent{0.0017} \\
\bottomrule
\end{tabular}
\begin{tablenotes}
\small
    \item [] (Percentage out of files and functions with floating point keywords.)
\item [a] GNU Scientific Library common mathematical functions
\item [b] GNU Scientific Library special functions
\item [c] GNU Scientific Library statistics functions
\end{tablenotes}
\end{threeparttable}

%% file: sections/case_study.tex
\section{Case Study: A Challenge Set of C Floating-Point Benchmarks}\label{sec:case-study}

To demonstrate how our dataset can support practical benchmarking, we construct a set of 59 challenge benchmarks in~C, which we upload along with the dataset.
The choice to extract benchmarks in C is arbitrary (from the list of most frequently used languages for floating-point code).
This case study is not intended as a definitive benchmark suite, but rather as an illustration of how our corpus can be used to generate realistic, ready-to-run benchmarks that reflect real-world floating-point practices.
Each benchmark is a self-contained file containing a function extracted from our corpus together with all necessary dependencies to compile and run it independently.

\paragraph{Sampling and Extraction}

The extraction process is semi-automated.
We uniformly sample 200 functions from projects using permissive licenses (MIT, BSD, Apache-2.0).
For each sampled function, we parse the source with Clang's Rust bindings\footnote{The extraction process is available as a subcommand in our framework.} and recursively constructs its dependency graph by traversing the project's directory structure.
Dependencies include type definitions and auxiliary functions.
When a new file is discovered, we also store the include directives and macros it contains.
When all dependencies are identified, we emit a single self-contained C file that aggregates the collected includes and macros, followed by the function and its dependencies in topological order.
Because we do not rely on build systems such as \texttt{make}, extraction can fail in the presence of conditional compilation or cyclic dependencies (46~out of~200~sampled functions).
Manual curation removes redundant dependencies and macros. To preserve self-containment and portability, we also exclude benchmarks that depend on external or platform-specific libraries (\eg{} OpenMP, CPython, \texttt{windows.h}), discarding 60 out of 155 remaining files.
We further discard duplicates\footnote{Although files in our dataset are deduplicated, identical function definitions may appear in distinct files, often due to code generation.} (34 out of  94) and empty functions (1 out of 94), resulting in a final set of 59 unique benchmarks.
We validate each extracted benchmark by compiling with both GCC 13.3.0 on Linux and Clang 21.1.8 on macOS.

\begin{figure*}[tb]
    \centering
    \begin{minipage}[t]{0.48\linewidth}
\begin{lstlisting}[style=highlight,language=C]
double calSum(double data[], int N, int I) {
    if (I == 0) { return data[0]; }
    
    double sum = 0.f;
    int idx = 0;

    if (I >= N) { idx = N; }
    else { idx = I; }

    for (int i = 0; i < idx; i++) {
        sum += data[I - i];
    }
    
    return sum;
}
\end{lstlisting}
    \end{minipage}
    \begin{minipage}[t]{0.48\linewidth}
\begin{lstlisting}[style=highlight,language=C,emph={calSum},emphstyle=\textbf]
void calAR(double PRICE[][4], int count, int N, double AR[]) {
    double hoValue[count];
    double olValue[count];

    for (int i = 0; i < count; i++) {
        hoValue[i] = PRICE[i][0] - PRICE[i][2];
        olValue[i] = PRICE[i][2] - PRICE[i][1];
    }

    for (int i = 0; i < count; i++) {
        double sum1 = calSum(hoValue, N, i);
        double sum2 = calSum(olValue, N, i);
        AR[i] = sum1 / sum2 * 100;
    }
}
\end{lstlisting}
    \end{minipage}
    \caption{
        Example C floating-point benchmark extracted from our corpus. Unlike FPBench procedures, \code{calAR} is not self-contained and calls \code{calSum} \cite{CalAR}.
    }
    \label{fig:ex-call}
    \Description{Two C functions, calSum and calAR, where calAR calls calSum to compute the sum of the first I elements of an array.}
\end{figure*}

\paragraph{Characteristics}

The resulting programs span multiple domains, including scientific computing (\Pcent{44}), embedded systems (\Pcent{12}), software toolchains (\Pcent{12}), graphics (\Pcent{10}), and educational material (\Pcent{10}). 
Their size ranges from~4~to~1021~lines of code (LoC), with a median of~30~LoC (against 3 for FPBench).
Unlike FPBench isolated numerical routines, the extracted functions preserve their original program context: several depend on auxiliary functions (\Pcent{29}) or user-defined structures and unions (\Pcent{37}), and most mix floating-point parameters with other types (\Pcent{66}).

\Cref{fig:ex-call} illustrates one such benchmark, where the target function~\code{calAR} calls an auxiliary routine~\code{calSum} that computes the sum of the first~\code{I} elements of an array. Benchmarks of this kind provide realistic cases for evaluating the interprocedural reasoning capabilities of floating-point analysis tools.

The extracted programs also differ from FPBench in the language constructs they employ.
A large fraction use macros (\Pcent{31}), pointer manipulation (\Pcent{63}), and explicit type casts (\Pcent{41}). They further exhibit floating-point usage patterns that extend beyond pure arithmetic or mathematical functions.
For example, \Cref{fig:ex-payload} shows a benchmark performing low-level memory manipulation to encode the least significant bits of an input into a NaN payload, while other programs use floating-point variables as loop counters.
These patterns illustrate how, due to C's permissive semantics, floating-point computations can appear in unexpected contexts.
By exposing such patterns, these benchmarks prompt tool designers to make explicit and informed choices about which language features to support and which to explicitly exclude by design.

\begin{wrapfigure}[14]{r}{.41\textwidth}
    \centering
\begin{lstlisting}[style=highlight,language=C]
static float quiet_nanf(float x) {
  uint32_t tmp;
  memcpy(&tmp, &x, 4);
  tmp |= 0x7fc00000lu;
  memcpy(&x, &tmp, 4);
  return x;
}
\end{lstlisting}
    \caption{ C floating-point benchmark extracted from our corpus that uses low-level memory manipulation to store the least significant bits of the input into a NaN value \cite{quietNaN}.}
    \label{fig:ex-payload}
    \Description{C function creating a quiet NaN by copying float bits to an integer, setting the NaN payload, and copying back.}
\end{wrapfigure}

By preserving real-world coding idioms that existing floating-point tools struggle to support, benchmarks extracted from our corpus will inevitably be challenging.
We envision such benchmarks to serve as a challenging and realistic target for future tool development.

\begin{table}[tb]
    \caption{Overview of representative freely available and recent floating-point reasoning tools. All listed tools
    support floating-point arithmetic.}
    \begin{adjustbox}{width=0.97\textwidth}
    \begin{threeparttable}[t]
    \centering
    \small 
    \begin{tabular}{@{}lccc@{}}
\toprule
Tool        & input lang. & aim                             & supported features \\
 \midrule 
 \emph{Static tools} & & & \\
\midrule
FPTaylor~\cite{solovyevRigorousEstimationFloatingPoint2018}
    & custom    & worst-case rounding error bounds & transcendental fncs. \\
\rowcolor{black!5}PRECiSA~\cite{titoloRigorousFloatingPointRoundOff2025}
    &  PVS      & \begin{tabular}[c]{@{}c@{}}worst-case rounding error bounds,\\ mixed-precision tuning\end{tabular} & \begin{tabular}[c]{@{}c@{}}transcendental fncs., \\ map and fold over lists, \\ non-recursive func. calls\tnote{a} \\ conditional stmt, loops\tnote{a} \end{tabular} \\
Daisy~\cite{abbasiModularOptimization2023,isychevScalingRoundingAnalysisFunctionalDS2023}
    & Scala     & worst-case rounding error bounds
    & \begin{tabular}[c]{@{}c@{}}transcendental fncs., \\ some vector and matrix operations, \\ non-recursive func. calls\end{tabular} \\
\rowcolor{black!5}Gappa~\cite{daumasM10Gappa}
    &  custom   & worst-case rounding error bounds & 11 rounding modes (only arithm.)\\
Frama-C~\cite{boldoFormalVerificationNumerical2011}
    & C & deductive program verification &
    \begin{tabular}[c]{@{}c@{}}essentially full C,\\ limited support for rounding errors via Gappa \end{tabular}\\
\rowcolor{black!5}KeY~\cite{abbasiCombiningRuleSMTbased2023}
    & Java & deductive program verification &
    \begin{tabular}[c]{@{}c@{}} subset of Java (roughly basic Java 1.2),\\ no autom. support for rounding errors \end{tabular}\\
Stainless~\cite{gilot2026floatStainlessTACAS}
    & Scala & deductive program verification &
    \begin{tabular}[c]{@{}c@{}}subset of Scala (mostly functional),\\ no autom. support for rounding errors \end{tabular}\\
\midrule
\emph{Dynamic tools} & & & \\
\midrule
\rowcolor{black!5}Herbie~\cite{panchekhaAutomaticallyImprovingAccuracy2015}
    & S-expressions & improve accuracy via rewriting &
    \begin{tabular}[c]{@{}c@{}}conditionals, \\fixed set of mathematical functions\end{tabular} \\
Verrou~\cite{fevotteStudyingNumericalQuality2017}
    & \begin{tabular}[c]{@{}c@{}}integrated \\to Valgrind\end{tabular}
    & \begin{tabular}[c]{@{}c@{}}numerical instability \\ via Monte Carlo arithmetic \end{tabular}&
    presumably all \\
\rowcolor{black!5}Verificarlo~\cite{verificarlo2016}
    & C, C++, Fortran
    & \begin{tabular}[c]{@{}c@{}}numerical instability \\ via Monte Carlo arithmetic \end{tabular}&
    presumably all \\
FPLearner~\cite{WangR24FPLearner}
    & C/C++ & mixed-precision tuning & presumably all\\
\rowcolor{black!5}NSan~\cite{courbetNSanFloatingpointNumerical2021}
    & LLVM &
    \begin{tabular}[c]{@{}c@{}}large rounding errors\\ via shadow execution\end{tabular}&
    presumably all \\
\bottomrule
\end{tabular}\label{tab:overview-tools}
\begin{tablenotes}
\small
\item [a] Not evaluated systematically in publications as far as we know.
\end{tablenotes}
\end{threeparttable}
\end{adjustbox}
\end{table}

\paragraph{Applicability to Existing Tools}

\Cref{tab:overview-tools} provides an overview of a selection of existing tools.
This list is not and is not meant to be exhaustive, rather we focus on available
and recent tools that could potentially benefit from a benchmark suite such as
the one that we extracted. (As stated previously, we do not intend our extracted
C benchmarks as a definitive benchmark suite.)
We focus on recent tools, i.e. tools whose code was updated in some way within
the last 5 years, as older tools tend to be difficult to run in practice. The
threshold of 5 years is arbitrary.

Many dynamic analysis tools target C/C++ in some way and so we expect
that they can be immediately run on the extracted benchmarks. To use the tool Herbie,
one would need to extract the numerical calculations out of the benchmarks.

To demonstrate that our benchmarks can be used in practice, we run NSan and Frama-C with Gappa as frontend on them.
In contrast to other tools, NSan requires an explicit entry point with a \code{main} function. 
To accommodate this requirement, we add a simple \code{main} function to each benchmark that invokes the extracted function with
fixed input values chosen to avoid runtime errors. 
Frama-C does not require an entry point and we therefore run it directly on the extracted benchmarks.
NSan successfully analyses all the benchmarks and does not report precision-loss warnings (this is not very surprising, as such warnings are typically highly input dependent).
Frama-C produces verification conditions for 54 out of the 59 benchmarks; the remaining five benchmarks use complex numbers, which are not supported by Frama-C.

For the remainder of the static analysis tools, one would need to translate
them from C to their respective input language. We expect that as part of this rewrite,
some of the currently unsupported features of each tool (\eg{} pointers) could also be expressed
in a way that the tool can handle. This effort is beyond the scope of this paper.

An alternative option for these tools is to extract programs from GitHub
directly written in the input language of interest.
The newly added support for floating-point arithmetic in
Stainless~\cite{gilot2026floatStainlessTACAS} successfully used Scala programs extracted in this code study for evaluation.

%% file: sections/discussion.tex
\section{Discussion and Conclusions}


\paragraph{\textbf{Benchmarks used in the literature reflect the current capabilities of tools, rather than the way developers actually write code}}
Although function sizes are comparable, our study shows that the internal structure and feature usage of FPBench functions differ from user-written code.
In particular, transcendental functions are overrepresented, whereas conditionals and their combinations with other control-flow constructs are significantly underrepresented.
FPBench thus aligns more closely with the capabilities of existing static floating-point reasoning techniques, which provide limited support (if any) for conditionals and loops
\footnote{Handling conditionals in static analyses of
\eg{} rounding errors is challenging, because the real-valued specification and
floating-point execution can take different paths through a program. Since
worst-case rounding error bounds grow with every loop iteration, it is unclear
how to find and prove loop invariants.}, and whose techniques are not modular.
Our study further reveals that, despite being frequently used for benchmarking in the literature, calls to GSL functions appear rarely (less than \Pcent{0.2}) at both the file and function level.

We believe that our corpus can be used to extract code for different benchmark sets.
Future research can also rely on this study to better gauge whether the complexity in a benchmark accurately reflects real-world floating-point usage.

\paragraph{\textbf{To be effective on user-written code, floating-point analysis tools must prioritize modularity and reasoning about branching behaviour.}}
Our study shows that floating-point code is typically modular, with most function bodies containing multiple calls to others. 
Conditionals are common, and loops almost always co-occur with conditionals. 
These characteristics suggest that the main challenge for analysis tools is not handling large, monolithic numerical kernels~\cite{dasScalableFloatErrorAnalysis2020}, but effectively analysing many small functions that combine modularity with non-trivial control flow.
While tools such as PRECISA 4.0~\cite{titoloRigorousFloatingPointRoundOff2025} and Daisy~\cite{abbasiModularOptimization2023} demonstrate promising approaches in this direction,
the majority of state-of-the-art roundoff error analysers target self-contained code fragments.

\paragraph{\textbf{Keyword-based approaches are unlikely to be effective for identifying floating-point code in dynamically typed languages.}}
This study presents part of the big picture, and extending the analysis to dynamically typed languages would provide a natural complement.
However, we believe this cannot be achieved by simply adapting our current methodology. 
Our findings show that in statically typed code, non-type keywords appear only rarely (\Pcent{6}--\Pcent{17} of functions depending on the language), 
and GSL-related keywords appear even less frequently, making non-standard math libraries an unreliable proxy.
Without explicit type annotations, a keyword-based approach in dynamically typed languages would therefore raise serious concerns about the validity of the results.
The presence of arithmetic operators would also be insufficient to identify floating-point usage, since they are commonly overloaded for integers or non-numerical data.
Obtaining reliable type information in such languages would require instrumenting and executing the code with representative inputs, 
which in turn demands resolving dependencies and reproducing realistic program executions (including GUI interactions where relevant). 
Designing methods to gather such information consistently and at scale remains an open challenge left for future work.

\paragraph{\textbf{Semantically characterising floating-point code in a quantitative, scalable, and automated fashion raises substantial challenges.}}
While it would be very interesting to identify a repository's application domain(s), detect common numerical algorithms, or assess numerical stability, these characterisations typically require manual annotation with multiple annotators to yield meaningful quantitative claims---we are not aware of automated tools that can do this reliably.
Type-based analyses are similarly not as easily automated since they generally require successfully building projects. Prior work has attempted this for Java using build tools such as Maven or Gradle~\cite{sulirQuantitativeStudyJava2016, hassanAutomaticBuildingJava2017, sulirLargeScaleDatasetLocal2020}, but reported low success rates, which can bias the analysis.
The process described in \Cref{sec:case-study} to compile C benchmarks required several weeks of work for a single programming language and required manual intervention for a large number of projects.
Developing a pipeline that addresses these issues is left for future work and can build on the baseline established by this paper.